\documentstyle[11pt,epic,eepic,epsf]{article}
\title{XXZ chain with a boundary}

\author{
Michio Jimbo\thanks{Department of Mathematics, Faculty of Science,
                            Kyoto University, Kyoto 606, Japan.},
Rinat Kedem\thanks{Research Institute for Mathematical Sciences,
                            Kyoto University, Kyoto 606, Japan.},
Takeo Kojima$^\dagger$, \cr
Hitoshi Konno\thanks{Yukawa Institute for Theoretical Physics,
                            Kyoto University, Kyoto 606, Japan.} and
Tetsuji Miwa$^\dagger$ \cr}

\date{November 1994}

\begin{document}

\maketitle
\newfont{\g}{eufm9}

\renewcommand{\theequation}{\thesection.\arabic{equation}}
\def\Phim#1{\mathrel{\mathop{\kern0pt \Phi}\limits^#1}}
\def\Psim#1{\mathrel{\mathop{\kern0pt \Psi^*}\limits^#1}}
\def\C{{\bf C}}
\def\Z{{\bf Z}}
\def\H{{\cal H}}
\def\T{{\cal T}}
\def\vep{\varepsilon}
\def\z{\zeta}
\def\mod{{\rm mod}~}
\def\glth{\widehat{gl}_2}
\def\slth{\widehat{}_2}
\def\uq{U_q(\slth)}				
\def\End{{\rm End}\,}
\def\R{{\cal R}}
\def\pr#1{(#1)_\infty}
\def\disp#1{{\displaystyle #1}}
\def\be{\begin{equation}}
\def\en{\end{equation}}
\def\goth#1{{#1}}
\def\displ#1{{\displaystyle #1}}
\def\n{\nonumber}
\def\bea{\begin{eqnarray}}
\def\ena{\end{eqnarray}}
\def\bean{\begin{eqnarray*}}
\def\enan{\end{eqnarray*}}
\def\refeq#1{(\ref{eqn:#1})}
\def\refto#1{\cite{#1}}
\def\lb#1{\label{eqn:#1}}
\def\rf#1{(\ref{eqn:#1})}
\def\disp{\displaystyle}
\def\Remark{\medskip\noindent {\sl Remark}\quad}
%
\def\C{{\bf C}}
\def\R{{\bf R}}
\def\Q{{\bf Q}}
\def\sp{\small{\bf p}}
\def\Ca{{\cal C}}  
\def\F{{\cal F}}
\def\la{\lambda}
\def\La{\Lambda}
\def\L{\Lambda}
\def\Ga{\Gamma}
\def\Th{\Theta}
\def\th{\theta}
\def\e{\epsilon}
\def\s{\sigma}
\def\x{\xi}


\def\bra#1{\langle #1 |}	
\def\ket#1{| #1\rangle}		
\def\br#1{\langle #1 \rangle}	
\def\vac{|{\rm vac}\rangle}	
\def\dvac{\langle {\rm vac}|}	
\def\XXX{{X\hskip-2pt X\hskip-2pt X}}
\def\XXZ{{X\hskip-2pt X\hskip-2pt Z}}
\def\hzero{{H_{B,0}}}
\def\tskl{T_B^{\rm fin}}
\def\hca{h_c^{(1)}}
\def\hcb{h_c^{(2)}}
\def\hfin{H_B^{\rm fin}}
\def\path{|p\rangle}
\def\Hom{{\rm Hom}\,}
\def\exp{{\rm exp}\,}
\def\ip{(q^4;q^4)_\infty}
\def\id{{\rm id}\,}
\def\tr{{\rm tr}\,}
\def\dim{{\rm dim}\,}
\def\ch{{\rm ch}\,}
\def\ad{{\rm ad}\,}
\def\wt{{\rm wt}\,}
\def\slt{\goth{sl}_2}				
\def\slth{\widehat{\goth{sl}}_2\hskip 1pt}	
\def\Uq{U_q(\goth{g})}				
\def\uq{U_q(\slth)}				
\def\goto#1{{\buildrel #1 \over \longrightarrow}}
\def\pr#1{\left(#1\right)_\infty}  
\def\prq#1{(#1;q^4)_\infty}  
\def\Rb{\overline{R}}  
\def\oR{{\overline R}} 
\def\Rc{\check{R}}  
\def\Rh{\widehat{R}} 
\def\Phit{\widetilde{\Phi}}  
\def\Phih{\widehat{\Phi}}    
\def\Psit{\widetilde{\Psi}}  
\def\Psih{\widehat{\Psi}}    
\def\Phip{{\Phi^{(p)}}}
\def\Psip{{\Psi^{(p)}}}
\def\Phim#1{\mathrel{\mathop{\kern0pt \Phi}\limits^#1}}
\def\Psim#1{\mathrel{\mathop{\kern0pt \Psi}\limits^#1}}
\def\Phin#1{\mathrel{\mathop{\kern0pt \Phit}\limits^#1}}
\def\Psin#1{\mathrel{\mathop{\kern0pt \Psit}\limits^#1}}
\def\Vp{\Psi_+(z)}	
\def\Vm{\Psi_-(z)}	

\def\Vh{V^{(h)}}
\def\veph{\hat{\vep}}
\def\betah{\hat{\beta}}
\def\alphah{\hat{\alpha}}
\def\Imy{{\rm Im}\,}
\catcode`@=11
\def\citen#1{%
\if@filesw \immediate \write \@auxout {\string \citation {#1}}\fi
\@tempcntb\m@ne \let\@h@ld\relax \def\@citea{}%
\@for \@citeb:=#1\do {%
  \@ifundefined {b@\@citeb}%
    {\@h@ld\@citea\@tempcntb\m@ne{\bf ?}%
    \@warning {Citation `\@citeb ' on page \thepage \space undefined}}%
    {\@tempcnta\@tempcntb \advance\@tempcnta\@ne
    \setbox\z@\hbox\bgroup 
    \ifnum0<0\csname b@\@citeb \endcsname \relax
       \egroup \@tempcntb\number\csname b@\@citeb \endcsname \relax
       \else \egroup \@tempcntb\m@ne \fi
    \ifnum\@tempcnta=\@tempcntb 
       \ifx\@h@ld\relax 
          \edef \@h@ld{\@citea\csname b@\@citeb\endcsname}%
       \else 
          \edef\@h@ld{\hbox{--}\penalty\@highpenalty
            \csname b@\@citeb\endcsname}%
       \fi
    \else   
       \@h@ld\@citea\csname b@\@citeb \endcsname
       \let\@h@ld\relax
    \fi}%
 \def\@citea{,\penalty\@highpenalty}%
}\@h@ld}
\def\@citex[#1]#2{\@cite{\citen{#2}}{#1}}%
\def\@cite#1#2{\leavevmode\unskip
  \ifnum\lastpenalty=\z@\penalty\@highpenalty\fi
  \ [{\multiply\@highpenalty 3 #1
      \if@tempswa,\penalty\@highpenalty\ #2\fi 
    }]\spacefactor\@m}
\catcode`@=12

\bigskip
\centerline{\sl To the memory of Mr. Toyosaburo  Taniguchi}

\begin{abstract}
The $\XXZ$ spin chain with a boundary magnetic field $h$ is
considered, using the vertex operator approach to diagonalize the
Hamiltonian. We find explicit bosonic formulas for the two vacuum
vectors with zero particle content.  There are three distinct regions
when $h\geq0$, in which the structure of the vacuum states is
different. Excited states are given by the action of vertex operators
on the vacuum states. We derive the boundary $S$-matrix and present an
integral formula for the correlation functions. The boundary
magnetization exhibits boundary hysteresis. We also discuss the
rational limit, the $\XXX$ model.

\end{abstract}

\setcounter{section}{0}
\setcounter{equation}{0}
\section{Introduction}

In the standard treatment of quantum integrable systems, one starts
with a finite size system and imposes periodic boundary conditions, in
order to ensure the commutativity of the transfer matrix. Recently,
there has been increasing interest in exploring other possible boundary
conditions compatible with integrability.

For lattice models with non-periodic boundary conditions, the works on
the Ising model are perhaps among the earliest (see McCoy-Wu
\cite{MWa,MWb}, Bariev \cite{Bariev}).  One should also mention Gaudin's
work \cite{Gaudin} concerning the Bose gas with a
delta-function interaction and the $\XXZ$ model (see also \cite{Alc}).
A systematic approach to this problem was initiated by Sklyanin
\cite{Skl} in the framework of the algebraic Bethe ansatz. It is well
known that, in the periodic case, one can construct a commuting family
of transfer matrices from a solution $R(\zeta)$ of the Yang-Baxter
equation (YBE). Sklyanin showed that a similar construction is
possible with the aid of a solution $K(\zeta)$ to the boundary YBE
(also known as the reflection equation)
\begin{equation}\label{eqn:BYBE}
K_2(\z_2)~R_{21}(\z_1\z_2)~K_1(\z_1)~R_{12}(\z_1/\z_2)
{}~=~
R_{21}(\z_1/\z_2)~K_1(\z_1)~R_{12}(\z_1\z_2)~K_2(\z_2)~.
\end{equation}
Several groups have investigated the solutions of (\ref{eqn:BYBE})
along with the Bethe ansatz equations for the  models associated with them
\cite{KuS,MN1,MN2,GMN,dVGR,IK,HSFY}.
For a recent review and further references, see \cite{YB}.

The boundary YBE (\ref{eqn:BYBE}) was originally formulated by
Cherednik \cite{Cher}, as a factorization condition for scattering at
a boundary wall. In massive integrable field theories, Ghoshal and
Zamolodchikov \cite{GZ} discussed integrable massive deformations of
conformal field theories in the presence of boundaries \cite{Cardy},
and developed a bootstrap approach to them (see
also~\cite{FS,FK,Sasaki} for related works).

One model treated by Sklyanin's method is the $\XXZ$ spin chain with a
boundary magnetic field:
\begin{equation}\label{eqn:HBOUND}
H_B^{\rm bare}~=~-\frac{1}{2}\sum_{k=1}^{\infty}
\Bigl(\sigma^x_{k+1}\sigma^x_k+\sigma^y_{k+1}\sigma^y_k
+\Delta \sigma^z_{k+1}\sigma^z_k\Bigr)
+h\sigma^z_1~.
\end{equation}
Here, we consider the model in the limit of the semi-infinite chain,
in the ferromagnetic regime, $\Delta\le-1$. In this paper, we present
an alternative approach to the solution of this model, applying the
diagonalization scheme developed in \cite{DFJMN,JMMN} for the `bulk'
Hamiltonian
\begin{equation}\label{eqn:HBULK}
H^{\rm bare}~=~-{1\over2}~\sum_{k=-\infty}^\infty~
\left(\sigma^x_{k+1}\sigma^x_{k}+\sigma^y_{k+1}\sigma^y_{k}
+\Delta\sigma^z_{k+1}\sigma^z_{k}\right).
\end{equation}
For a systematic account of this method, see \cite{JM}.

The ground state energy of the bare Hamiltonian~\refeq{HBOUND} is a
divergent scalar, $c(\Delta,h)$, which we subtract. We subsequently
consider the spectrum of the `renormalized Hamiltonian,'
\be H_B~=~H_B^{\rm bare}-c(\Delta,h)~,\label{RENOR} \en
which has a ground state energy equal to 0.

The starting point in the bulk theory is to identify the space of
eigenvectors of the Hamiltonian (\ref{RENOR}) (`space of states') with
the tensor product ${\cal H}\otimes{\cal H}^*$, where ${\cal H}={\cal
H}^{(0)}\oplus{\cal H}^{(1)}$ is the direct sum of level one
integrable modules of $\uq$, and ${\cal H}^*$ its dual.  Here,
\be
\Delta~=~{q+q^{-1}\over2}~,\label{delta}
\en
and $-1< q <0$.

In contrast, we relate the space of states of the boundary Hamiltonian
(\ref{eqn:HBOUND}) to the half space ${\cal H}$. Accordingly, we
rewrite Sklyanin's transfer matrix $T_B(\zeta)$, on the semi-infinite
lattice, in terms of the type I vertex operators $\Phi_\vep(\zeta)$
(see (\ref{eqn:TB}) below).  The renormalized Hamiltonian
(\ref{RENOR}) is defined as the derivative of $T_B(\zeta)$ at
$\zeta=1$.

It should be emphasized that, unlike in the bulk theory, the algebra
$\uq$ does {\it not} commute with the boundary Hamiltonian
(\ref{eqn:HBOUND}), and  hence is not a symmetry algebra of this model.
Nevertheless, we will show that one can use ${\cal H}$ and
$\Phi_\vep(\z)$ to describe the space of states.  A simple
bosonization formula is available for the spaces ${\cal H}^{(i)}$ and
the vertex operators \cite{JMMN,FJ}. Our basic result here is that, in
the bosonization language, the vacuum vectors in each sector ${\cal
H}^{(i)}$ have the remarkably simple form:
\begin{equation}\label{eqn:grst}
\ket{i}_B~=~e^{F_i}~\ket{i}~,
\end{equation}
where $\ket{i}$ is the highest weight vector of ${\cal H}^{(i)}$, and
$F_i$ is quadratic in the bosonic operators (see (\ref{eqn:Fi})).
This fact allows us to compute the spectrum of (\ref{RENOR}), and the
spin correlation functions. We subsequently find the following features.

In the presence of the boundary term, the Hamiltonian
(\ref{eqn:HBOUND}) lacks spin-reversal symmetry, and the two vacuum
vectors, $|i\rangle_B$, carry different energies. This energy difference
(see (\ref{eqn:e1})) corresponds to the binding energy of the boundary
bound state of \cite{GZ}.

Although the expression \refeq{grst} makes sense in the bosonic Fock
space, there is no {\it a priori} guarantee that $\ket{i}_B$ are
well defined
for all values of $h$.  We surmise that they exist as
long as they are regular at $q=0$. According to this criterion,
$\ket{1}_B$ exists if and only if $h<h_c^{(1)}$ or $h>h_c^{(2)}$, and
$\ket{0}_B$ exists if and only if $h>-h_c^{(1)}$ or $h<-h_c^{(2)}$.
Here, the critical fields are
\begin{equation}\label{eqn:hcrit}
 h_c^{(1)}~ =~
\frac{(1+q)^2}{-4q}=-\frac{\Delta+1}{2}, \qquad h_c^{(2)} ~=~
\frac{(1-q)^2}{-4q}=- \frac{\Delta-1}{2}~.
\end{equation}
We will show, in Section 3, that our criterion for existence
agrees with the singularity structure of the boundary $S$-matrix.

As in the bulk theory, excited states are created by action of
type II vertex operators $\Psi^*_\mu(\xi)$ on $\ket{i}_B$.
Using our expression for the vacuum states (\ref{eqn:grst}), we deduce
that single--particle states obey the relations
\[
\Psi^*_\mu(\xi)\ket{i}_B~=~
M^{(i)\mu}_\mu(\xi)~\Psi^*_\mu(\xi^{-1})~\ket{i}_B~,
\qquad (\mu=\pm)~.
\]
The matrix $M^{(i)}(\xi)$ is therefore the boundary $S$-matrix,
analogous to that formulated in \cite{GZ} for integrable field
theories.

The bosonic expression for the vacuum vectors enables us to derive
integral formulas for the spin correlation functions.  The boundary
magnetization is an especially simple case of these correlation
functions (see (\ref{eqn:Mag})). For example,
\[
-~\frac{{}_B\bra{0}\sigma^z_1\ket{0}_B}{{}_B\langle 0|0\rangle_B}
{}~=~1+2~\sum_{l=1}^{\infty}~\frac{(-q^{2})^l(1-r)^2}{ (1-q^{2l}r)^2}~,
\]
where
$$
h=\frac{1-q^2}{-4q}\frac{1+r}{1-r}~.
$$  The values of $h$ where
both vacuum states $\ket{i}_B$ coexist are those
where boundary hysteresis, or wetting, occurs~\cite{MWa,MWb,MW}.

The text is organized as follows.  In Section 2, we  briefly review
Sklyanin's theory, and formulate the vertex operator approach to the
boundary theory.
We calculate the energy difference between the two vacuum vectors
$|i\rangle_B$.  In
Section 3, we find explicit expressions for these vectors, by using
the bosonization formulas for vertex operators.
We then calculate the boundary $S$-matrix. Section 4 is devoted to
finding integral formulas for the spin correlation functions in
general, and the boundary magnetization in particular. The rational
limit, which corresponds to the $\XXX$ model, is discussed in Section
5. In Section 6 we discuss our results and note some open questions.


\setcounter{equation}{0}
\def\id{{\rm id}}

\section{Eigenvalues of the transfer matrix}

\subsection{Sklyanin's formulation}
In this section we recall Sklyanin's results in the context of the
$\XXZ$ chain on the finite lattice,
\begin{equation}\label{eqn:Hfin}
\hfin=-\frac{1}{2}\sum_{k=1}^{N-1}
\Bigl(\sigma^x_{k+1}\sigma^x_k+\sigma^y_{k+1}\sigma^y_k
+\Delta \sigma^z_{k+1}\sigma^z_k\Bigr)
+h_-\sigma^z_1+h_+\sigma^z_N~.
\end{equation}
This will motivate our construction of the transfer matrix for the
semi-infinite lattice below.

The transfer matrix is constructed as follows.  Below,
the reader is referred to Appendix A for notation. Let $R(\z)$ be the
$R$-matrix (\ref{eqn:R}) of the six--vertex model, where $\z$ is the
multiplicative spectral parameter. We also specify a matrix $K(\z)$,
corresponding to an interaction at the boundary, which satisfies the
boundary YBE (\ref{eqn:BYBE}).  Here, we consider only the diagonal
solution~\cite{Cher}
\begin{equation}\label{eqn:K}
K(\z)=K(\z;r)=
\frac{1}{f(\z;r)}
\left(\matrix{\displaystyle\frac{1-r\z^2}{\z^2-r} & 0\cr 0 & 1 \cr}\right).
\end{equation}
Nondiagonal solutions correspond to fields coupled to the other spin
components.

The scalar $f(\z;r)$ in (\ref{eqn:K}) is
\begin{equation}\label{eqn:f}
f(\z;r)=\frac{\varphi(\z^{-2};r)}{\varphi(\z^2;r)},
\qquad
\varphi(z;r)=\frac{\pr{q^4rz;q^4}}{\pr{q^2rz;q^4}}
\frac{\pr{q^6z^2;q^8}}{\pr{q^8z^2;q^8}}~,
\end{equation}
and is chosen so that the $K$-matrix
obeys the
relations
\begin{eqnarray}
&&K(\z)K(\z^{-1})=1, \quad{\hbox{\rm (Boundary unitarity),}}
\label{eqn:bunitarity}\\
&&K^{b}_{a}(-q^{-1}\z^{-1})
=\sum_{a',b'}R^{-a~b}_{a'\,-b'}\bigl(-q\z^{2}\bigr)
K^{a'}_{b'}(\z), \quad{\hbox{\rm (Boundary crossing).}}
\label{eqn:bcrossing}
\end{eqnarray}
These ensure that the transfer matrix (\ref{eqn:TB}) satisfies the
unitarity and crossing relations (\ref{eqn:unitT}),
(\ref{eqn:crossT}).

\begin{figure}[htb]
\begin{center}
\setlength{\unitlength}{0.0125in}
\begin{picture}(330,356)(0,-10)
\drawline(315,45)(280,25)
\drawline(285.954,30.706)(280.000,25.000)(287.938,27.233)
\drawline(280,60)(315,45)
\drawline(306.859,46.313)(315.000,45.000)(308.435,49.990)
\drawline(155,250)(125,250)
\drawline(133.000,252.000)(125.000,250.000)(133.000,248.000)
\drawline(130,310)(155,310)
\drawline(147.000,308.000)(155.000,310.000)(147.000,312.000)
\drawline(60,320)(60,240)
\drawline(58.000,248.000)(60.000,240.000)(62.000,248.000)
\drawline(180,320)(180,240)
\drawline(178.000,248.000)(180.000,240.000)(182.000,248.000)
\drawline(220,320)(220,240)
\drawline(218.000,248.000)(220.000,240.000)(222.000,248.000)
\drawline(260,245)(275,260)
\drawline(260,265)(275,280)
\drawline(260,285)(275,300)
\drawline(260,305)(275,320)
\drawline(260,320)(260,240)
\drawline(5,240)(20,255)
\drawline(5,260)(20,275)
\drawline(5,280)(20,295)
\drawline(5,300)(20,315)
\drawline(20,320)(20,240)
\drawline(125,195)(125,115)
\drawline(123.000,123.000)(125.000,115.000)(127.000,123.000)
\drawline(165,155)(85,155)
\drawline(93.000,157.000)(85.000,155.000)(93.000,153.000)
\drawline(40,0)(55,15)
\drawline(40,20)(55,35)
\drawline(40,40)(55,55)
\drawline(40,60)(55,75)
\drawline(55,80)(55,0)
\drawline(90,20)(55,40)
\drawline(62.938,37.767)(55.000,40.000)(60.954,34.294)
\drawline(55,40)(90,55)
\drawline(83.435,50.010)(90.000,55.000)(81.859,53.687)
\drawline(315,85)(315,5)
\drawline(315,70)(330,85)
\drawline(315,50)(330,65)
\drawline(315,30)(330,45)
\drawline(315,10)(330,25)
\drawline(20,280)
	(23.750,278.770)
	(27.500,277.578)
	(31.250,276.426)
	(35.000,275.312)
	(38.750,274.238)
	(42.500,273.203)
	(46.250,272.207)
	(50.000,271.250)
	(53.750,270.332)
	(57.500,269.453)
	(61.250,268.613)
	(65.000,267.812)
	(68.750,267.051)
	(72.500,266.328)
	(76.250,265.645)
	(80.000,265.000)
	(83.750,264.395)
	(87.500,263.828)
	(91.250,263.301)
	(95.000,262.812)
	(98.750,262.363)
	(102.500,261.953)
	(106.250,261.582)
	(110.000,261.250)
	(113.750,260.957)
	(117.500,260.703)
	(121.250,260.488)
	(125.000,260.312)
	(128.750,260.176)
	(132.500,260.078)
	(136.250,260.020)
	(140.000,260.000)
	(143.750,260.020)
	(147.500,260.078)
	(151.250,260.176)
	(155.000,260.312)
	(158.750,260.488)
	(162.500,260.703)
	(166.250,260.957)
	(170.000,261.250)
	(173.750,261.582)
	(177.500,261.953)
	(181.250,262.363)
	(185.000,262.812)
	(188.750,263.301)
	(192.500,263.828)
	(196.250,264.395)
	(200.000,265.000)
	(203.750,265.645)
	(207.500,266.328)
	(211.250,267.051)
	(215.000,267.812)
	(218.750,268.613)
	(222.500,269.453)
	(226.250,270.332)
	(230.000,271.250)
	(233.750,272.207)
	(237.500,273.203)
	(241.250,274.238)
	(245.000,275.312)
	(248.750,276.426)
	(252.500,277.578)
	(256.250,278.770)
	(260.000,280.000)

\drawline(20,280)
	(23.750,281.230)
	(27.500,282.422)
	(31.250,283.574)
	(35.000,284.688)
	(38.750,285.762)
	(42.500,286.797)
	(46.250,287.793)
	(50.000,288.750)
	(53.750,289.668)
	(57.500,290.547)
	(61.250,291.387)
	(65.000,292.188)
	(68.750,292.949)
	(72.500,293.672)
	(76.250,294.355)
	(80.000,295.000)
	(83.750,295.605)
	(87.500,296.172)
	(91.250,296.699)
	(95.000,297.188)
	(98.750,297.637)
	(102.500,298.047)
	(106.250,298.418)
	(110.000,298.750)
	(113.750,299.043)
	(117.500,299.297)
	(121.250,299.512)
	(125.000,299.688)
	(128.750,299.824)
	(132.500,299.922)
	(136.250,299.980)
	(140.000,300.000)
	(143.750,299.980)
	(147.500,299.922)
	(151.250,299.824)
	(155.000,299.688)
	(158.750,299.512)
	(162.500,299.297)
	(166.250,299.043)
	(170.000,298.750)
	(173.750,298.418)
	(177.500,298.047)
	(181.250,297.637)
	(185.000,297.188)
	(188.750,296.699)
	(192.500,296.172)
	(196.250,295.605)
	(200.000,295.000)
	(203.750,294.355)
	(207.500,293.672)
	(211.250,292.949)
	(215.000,292.188)
	(218.750,291.387)
	(222.500,290.547)
	(226.250,289.668)
	(230.000,288.750)
	(233.750,287.793)
	(237.500,286.797)
	(241.250,285.762)
	(245.000,284.688)
	(248.750,283.574)
	(252.500,282.422)
	(256.250,281.230)
	(260.000,280.000)

\put(250,20){\makebox(0,0)[lb]{\raisebox{0pt}[0pt][0pt]{\shortstack[l]{{\twlrm $\zeta^{-1}$}}}}}
\put(125,275){\makebox(0,0)[lb]{\raisebox{0pt}[0pt][0pt]{\shortstack[l]{{\twlrm $\cdots$}}}}}
\put(250,55){\makebox(0,0)[lb]{\raisebox{0pt}[0pt][0pt]{\shortstack[l]{{\twlrm $\zeta$}}}}}
\put(275,5){\makebox(0,0)[lb]{\raisebox{0pt}[0pt][0pt]{\shortstack[l]{{\twlrm $b$}}}}}
\put(275,70){\makebox(0,0)[lb]{\raisebox{0pt}[0pt][0pt]{\shortstack[l]{{\twlrm $a$}}}}}
\put(185,35){\makebox(0,0)[lb]{\raisebox{0pt}[0pt][0pt]{\shortstack[l]
{{\twlrm ${(K_-)}^a_b(\zeta)=$}}}}}
\put(-25,35){\makebox(0,0)[lb]{\raisebox{0pt}[0pt][0pt]{\shortstack[l]{{\twlrm ${(K_+)}^a_b(\zeta)=$}}}}}
\put(10,155){\makebox(0,0)[lb]{\raisebox{0pt}[0pt][0pt]{\shortstack[l]
{{\twlrm $R^{ab}_{cd}(\zeta)=$}}}}}
\put(50,330){\makebox(0,0)[lb]{\raisebox{0pt}[0pt][0pt]{\shortstack[l]{{\twlrm $N$}}}}}
\put(175,330){\makebox(0,0)[lb]{\raisebox{0pt}[0pt][0pt]{\shortstack[l]
{{\twlrm $2$}}}}}
\put(210,330){\makebox(0,0)[lb]{\raisebox{0pt}[0pt][0pt]{\shortstack[l]{{\twlrm $1$}}}}}
\put(125,225){\makebox(0,0)[lb]{\raisebox{0pt}[0pt][0pt]{\shortstack[l]{{\twlrm $\zeta^{-1}$}}}}}
\put(125,325){\makebox(0,0)[lb]{\raisebox{0pt}[0pt][0pt]{\shortstack[l]{{\twlrm $\zeta$}}}}}
\put(115,200){\makebox(0,0)[lb]{\raisebox{0pt}[0pt][0pt]{\shortstack[l]{{\twlrm $a$}}}}}
\put(175,155){\makebox(0,0)[lb]{\raisebox{0pt}[0pt][0pt]{\shortstack[l]{{\twlrm $b$}}}}}
\put(115,95){\makebox(0,0)[lb]{\raisebox{0pt}[0pt][0pt]{\shortstack[l]{{\twlrm $c$}}}}}
\put(65,155){\makebox(0,0)[lb]{\raisebox{0pt}[0pt][0pt]{\shortstack[l]{{\twlrm $d$}}}}}
\put(135,165){\makebox(0,0)[lb]{\raisebox{0pt}[0pt][0pt]{\shortstack[l]{{\twlrm $\zeta$}}}}}
\put(75,60){\makebox(0,0)[lb]{\raisebox{0pt}[0pt][0pt]{\shortstack[l]{{\twlrm $b$}}}}}
\put(75,5){\makebox(0,0)[lb]{\raisebox{0pt}[0pt][0pt]{\shortstack[l]{{\twlrm $a$}}}}}
\put(100,45){\makebox(0,0)[lb]{\raisebox{0pt}[0pt][0pt]{\shortstack[l]{{\twlrm $\zeta$}}}}}
\put(100,15){\makebox(0,0)[lb]{\raisebox{0pt}[0pt][0pt]{\shortstack[l]{{\twlrm $\zeta^{-1}$}}}}}
\end{picture}
\caption{Transfer matrix on a finite chain. The spectral parameters attached
to the lines $1,\cdots,N$ are chosen to be $1$.}
\label{f1}
\end{center}
\end{figure}
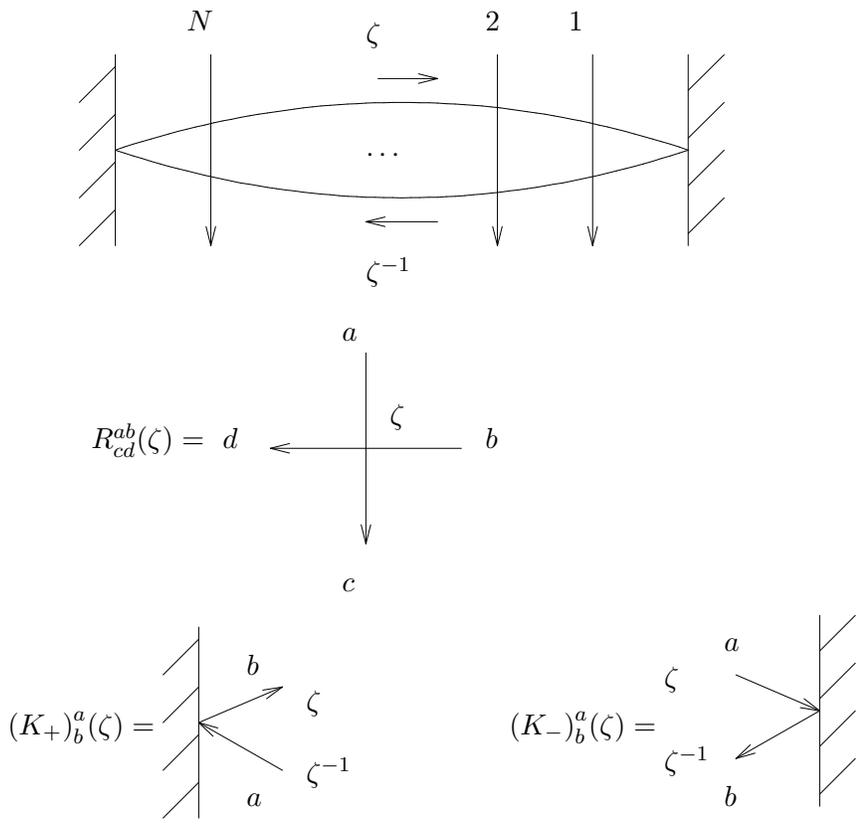

A commuting family of transfer matrices is constructed from $R$ and
$K$ as follows~\cite{Skl}:
\begin{equation}\label{eqn:TBfin}
\tskl(\z)=\tr_{V_0}\Bigl(K_+(\z)\T(\z^{-1})^{-1}K_-(\z)\T(\z)\Bigr).
\end{equation}
Here
\[
\T(\z)=R_{01}(\z)\cdots R_{0N}(\z) ~\in~
\End\left(V_0\otimes V_1\otimes\cdots\otimes V_N\right)
\]
denotes the monodromy matrix, $V_j$ are copies of $\C^2$, and
\[
K_-(\z)=K(\z;r_-),~~ K_+(\z)=K(-q^{-1}\z^{-1};r_+)^t
{}~\in\End\Bigl(V_0\Bigr)~,
\]
where $r_\pm$ are arbitrary parameters.  Graphically, the transfer
matrix (\ref{eqn:TBfin}) is represented in Figure~\ref{f1}.

With these definitions, the following statements hold \cite{Skl}:
\begin{enumerate}
\item The transfer matrices (\ref{eqn:TBfin}) form a commutative family:
\[
[\tskl(\z),\tskl(\z')]=0\qquad  \forall \z,\z'.
\]
\item
The Hamiltonian (\ref{eqn:Hfin}) is obtained as
\begin{equation}\label{eqn:HviaT}
\frac{d}{d\z}\tskl(\z)\Bigl|_{\z=1}
=\frac{4q}{1-q^2}\hfin+{\rm const}.
\end{equation}
where $\Delta$ is defined by (\ref{delta}) and
\be
h_{\pm}=\mp\frac{1-q^2}{-4q}\frac{1+r_\pm}{1-r_\pm}~.
\en
\end{enumerate}
In \cite{Skl}, an algebraic Bethe ansatz is constructed for the
eigenvectors of the transfer matrix (\ref{eqn:TBfin}).  The Bethe
ansatz equations for (\ref{eqn:Hfin}) were previously derived in
\cite{Gaudin,Alc} using the coordinate Bethe ansatz.

\subsection{The semi-infinite spin chain}
We now consider the $\XXZ$ Hamiltonian in the limit of the
semi-infinite chain, (\ref{RENOR}), with $h=h_-$ and $r=r_-$.
Since, under conjugation of $H_B$ by the spin-reversal
operator $\prod \sigma^x_j$, the sign of the boundary term is
reversed, we can restrict our discussion to $h\ge0$, or
\[
-1\le r \le 1.
\]
The free boundary condition, $h=0$, is $r=-1$, whereas the fixed
boundary condition,
$h=\infty$, is $r=1$.  When $r=0$ (or $r=\infty$),
the Hamiltonian (\ref{eqn:HBOUND}) formally enjoys
$U_q(sl_2)$--invariance~\cite{PS}.

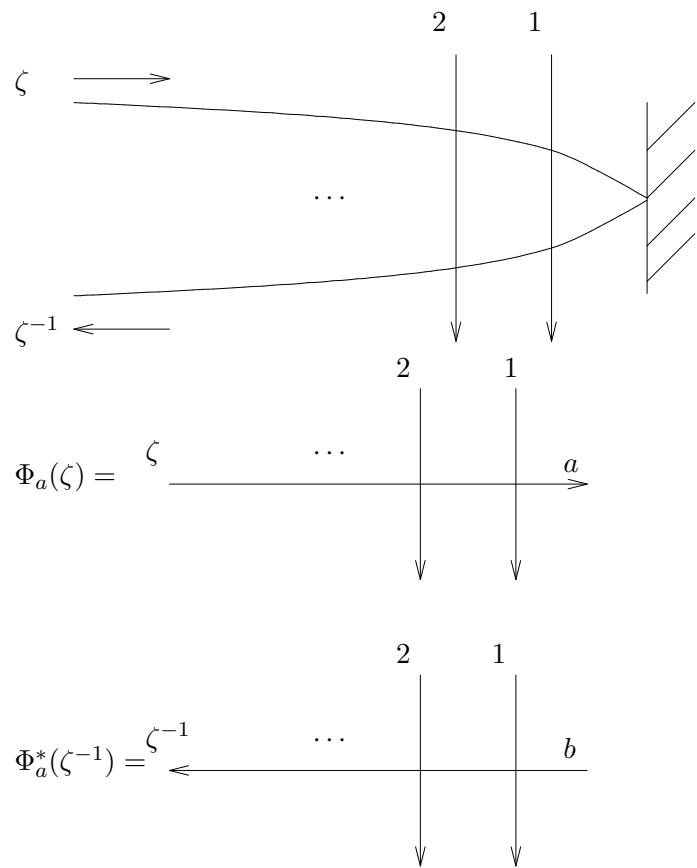
\begin{figure}[htb]
\begin{center}
\setlength{\unitlength}{0.0125in}
\begin{picture}(285,376)(0,-10)
\drawline(170,80)(170,0)
\drawline(168.000,8.000)(170.000,0.000)(172.000,8.000)
\drawline(210,80)(210,0)
\drawline(208.000,8.000)(210.000,0.000)(212.000,8.000)
\drawline(170,200)(170,120)
\drawline(168.000,128.000)(170.000,120.000)(172.000,128.000)
\drawline(210,200)(210,120)
\drawline(208.000,128.000)(210.000,120.000)(212.000,128.000)
\drawline(240,40)(65,40)
\drawline(73.000,42.000)(65.000,40.000)(73.000,38.000)
\drawline(65,160)(240,160)
\drawline(232.000,158.000)(240.000,160.000)(232.000,162.000)
\drawline(65,225)(25,225)
\drawline(33.000,227.000)(25.000,225.000)(33.000,223.000)
\drawline(25,330)(65,330)
\drawline(57.000,328.000)(65.000,330.000)(57.000,332.000)
\drawline(185,340)(185,220)
\drawline(183.000,228.000)(185.000,220.000)(187.000,228.000)
\drawline(225,340)(225,220)
\drawline(223.000,228.000)(225.000,220.000)(227.000,228.000)
\drawline(285,265)(265,245)
\drawline(285,280)(265,260)
\drawline(285,300)(265,280)
\drawline(285,320)(265,300)
\drawline(265,320)(265,240)
\drawline(25,239)	(27.700,239.119)
	(30.375,239.238)
	(33.027,239.357)
	(35.656,239.476)
	(38.260,239.595)
	(40.842,239.714)
	(43.400,239.833)
	(45.935,239.952)
	(48.448,240.071)
	(50.937,240.190)
	(55.848,240.429)
	(60.670,240.669)
	(65.404,240.909)
	(70.051,241.150)
	(74.612,241.392)
	(79.087,241.635)
	(83.480,241.879)
	(87.789,242.125)
	(92.018,242.372)
	(96.166,242.620)
	(100.235,242.870)
	(104.226,243.121)
	(108.141,243.375)
	(111.980,243.630)
	(115.744,243.888)
	(119.435,244.147)
	(123.054,244.409)
	(126.601,244.674)
	(130.079,244.941)
	(133.488,245.210)
	(136.830,245.483)
	(140.105,245.758)
	(143.314,246.036)
	(146.460,246.317)
	(149.542,246.602)
	(152.562,246.889)
	(155.522,247.181)
	(158.422,247.475)
	(161.264,247.774)
	(164.048,248.076)
	(166.776,248.382)
	(169.449,248.692)
	(172.068,249.006)
	(174.635,249.325)
	(177.150,249.648)
	(182.029,250.307)
	(186.716,250.985)
	(191.219,251.683)
	(195.548,252.401)
	(199.711,253.141)
	(203.719,253.903)
	(207.580,254.689)
	(211.303,255.499)
	(214.897,256.334)
	(218.372,257.195)
	(221.736,258.084)
	(225.000,259.000)

\drawline(225,259)	(227.993,259.969)
	(231.325,261.258)
	(235.113,262.926)
	(239.475,265.031)
	(241.907,266.266)
	(244.527,267.633)
	(247.349,269.138)
	(250.387,270.789)
	(253.657,272.593)
	(257.173,274.558)
	(260.949,276.692)
	(265.000,279.000)

\drawline(25,320)	(27.700,319.881)
	(30.375,319.762)
	(33.027,319.643)
	(35.656,319.524)
	(38.260,319.405)
	(40.842,319.286)
	(43.400,319.167)
	(45.935,319.048)
	(48.448,318.929)
	(50.937,318.810)
	(55.848,318.571)
	(60.670,318.331)
	(65.404,318.091)
	(70.051,317.850)
	(74.612,317.608)
	(79.087,317.365)
	(83.480,317.121)
	(87.789,316.875)
	(92.018,316.628)
	(96.166,316.380)
	(100.235,316.130)
	(104.226,315.879)
	(108.141,315.625)
	(111.980,315.370)
	(115.744,315.112)
	(119.435,314.853)
	(123.054,314.591)
	(126.601,314.326)
	(130.079,314.059)
	(133.488,313.790)
	(136.830,313.517)
	(140.105,313.242)
	(143.314,312.964)
	(146.460,312.683)
	(149.542,312.398)
	(152.562,312.111)
	(155.522,311.820)
	(158.422,311.525)
	(161.264,311.226)
	(164.048,310.924)
	(166.776,310.618)
	(169.449,310.308)
	(172.068,309.994)
	(174.635,309.675)
	(177.150,309.352)
	(182.029,308.693)
	(186.716,308.015)
	(191.219,307.317)
	(195.548,306.599)
	(199.711,305.859)
	(203.719,305.097)
	(207.580,304.311)
	(211.303,303.501)
	(214.897,302.666)
	(218.372,301.805)
	(221.736,300.916)
	(225.000,300.000)

\drawline(225,300)	(227.993,299.031)
	(231.325,297.742)
	(235.113,296.074)
	(239.475,293.969)
	(241.907,292.734)
	(244.527,291.367)
	(247.349,289.862)
	(250.387,288.211)
	(253.657,286.407)
	(257.173,284.442)
	(260.949,282.308)
	(265.000,280.000)

\put(160,85){\makebox(0,0)[lb]{\raisebox{0pt}[0pt][0pt]{\shortstack[l]{{\twlrm $2$}}}}}
\put(160,205){\makebox(0,0)[lb]{\raisebox{0pt}[0pt][0pt]{\shortstack[l]{{\twlrm $2$}}}}}
\put(200,85){\makebox(0,0)[lb]{\raisebox{0pt}[0pt][0pt]{\shortstack[l]{{\twlrm $1$}}}}}
\put(205,205){\makebox(0,0)[lb]{\raisebox{0pt}[0pt][0pt]{\shortstack[l]{{\twlrm $1$}}}}}
\put(55,50){\makebox(0,0)[lb]{\raisebox{0pt}[0pt][0pt]{\shortstack[l]{{\twlrm $\zeta^{-1}$}}}}}
\put(125,50){\makebox(0,0)[lb]{\raisebox{0pt}[0pt][0pt]{\shortstack[l]{{\twlrm $\cdots$}}}}}
\put(125,170){\makebox(0,0)[lb]{\raisebox{0pt}[0pt][0pt]{\shortstack[l]{{\twlrm $\cdots$}}}}}
\put(125,277){\makebox(0,0)[lb]{\raisebox{0pt}[0pt][0pt]{\shortstack[l]{{\twlrm $\cdots$}}}}}
\put(55,170){\makebox(0,0)[lb]{\raisebox{0pt}[0pt][0pt]{\shortstack[l]{{\twlrm $\zeta$}}}}}
\put(0,40){\makebox(0,0)[lb]{\raisebox{0pt}[0pt][0pt]{\shortstack[l]
{{\twlrm $\Phi^*_a(\zeta^{-1})=$}}}}}
\put(0,160){\makebox(0,0)[lb]{\raisebox{0pt}[0pt][0pt]{\shortstack[l]{{\twlrm $\Phi_a(\zeta)=$}}}}}
\put(230,165){\makebox(0,0)[lb]{\raisebox{0pt}[0pt][0pt]{\shortstack[l]{{\twlrm $a$}}}}}
\put(230,45){\makebox(0,0)[lb]{\raisebox{0pt}[0pt][0pt]{\shortstack[l]{{\twlrm $b$}}}}}
\put(0,220){\makebox(0,0)[lb]{\raisebox{0pt}[0pt][0pt]{\shortstack[l]{{\twlrm $\zeta^{-1}$}}}}}
\put(0,325){\makebox(0,0)[lb]{\raisebox{0pt}[0pt][0pt]{\shortstack[l]{{\twlrm $\zeta$}}}}}
\put(175,350){\makebox(0,0)[lb]{\raisebox{0pt}[0pt][0pt]{\shortstack[l]{{\twlrm $2$}}}}}
\put(215,350){\makebox(0,0)[lb]{\raisebox{0pt}[0pt][0pt]{\shortstack[l]{{\twlrm $1$}}}}}
\end{picture}
\caption{Transfer matrix on the semi-infinite chain. By convention,
the lattice sites are numbered $k=1,2,3,\cdots $ from right to left.
}
\label{f2}
\end{center}
\end{figure}

The transfer matrix $T_B(\z)$ corresponding to the limit of
(\ref{eqn:HBOUND}) is depicted in Figure~\ref{f2}. It describes a
semi-infinite two dimensional lattice, with alternating spectral
parameters.

In the na{\"\i}ve tensor product space $\cdots\otimes V\otimes
V\otimes V$, the eigenstates of $H_B$ which have finite eigenvalues
span a subspace $\H$, which we call the space of states. Much
insight about this space can be gained by examining the extreme
anisotropic limit, $q\rightarrow 0$.  In this limit, the Hamiltonian
$-2qH_B$ scales to
\be\lb{BH0}
\hzero~=~{1\over2}~\sum_{k=1}^\infty(\sigma^z_{k+1}
\sigma^z_k+1)+{1+r\over 2(1-r)}\sigma^z_1~,
\en
where a scalar term has been added to ensure that the lowest
eigenvalue of $\hzero$ is 0.

When $r=-1$, $\hzero$ has two antiferromagnetic
ground states, $|p^{(i)}\rangle$,
\be\lb{PATHVAC}
\ket{p^{(0)}}=\cdots \otimes v_+\otimes v_-\otimes v_+\otimes v_-,
\qquad
\ket{p^{(1)}}=\cdots \otimes v_-\otimes v_+\otimes v_-\otimes v_+.
\en
Accordingly, $\H$ splits into two subsectors, $\H^{(0)}$ and
$\H^{(1)}$, where $\H^{(i)}$ is the span of vectors
$|p\rangle=\otimes_{k=1}^\infty v_{p(k)}$, called paths,  labeled by maps
$p:\Z_{\ge1}\rightarrow\{\pm\}$ satisfying the asymptotic boundary condition
\be\lb{BCI}
p(k)=(-1)^{k+i}\hbox{ for }k\gg1.
\en

When $r\ne-1$, the eigenvalues of $\hzero$ corresponding to
$|p^{(0)}\rangle$ and $|p^{(1)}\rangle$ differ:
\[
\hzero\ket{p^{(0)}}=-{1+r\over2(1-r)}\,\ket{p^{(0)}}~,\quad
\hzero\ket{p^{(1)}}={1+r\over2(1-r)}\,\ket{p^{(1)}}~.
\]
Therefore, $\ket{p^{(1)}}$ is no longer a ground state.

\medskip

The splitting into two sectors $\H^{(i)}\,(i=0,1)$ described above
persists for general values of $r$, and $-1<q<0$.  For $q\ne0$, an
eigenvector of $H_B$ is an ``infinite linear combination'' of the
paths with one of the boundary conditions \refeq{BCI}, with $i=0,1$.
Near $r=-1$, each sector contains a state $\ket{i}_B$, unique up to a
scalar multiple, which has the lowest energy in that sector.  We call
$\ket{i}_B$ the vacuum vectors, and extend the definition to all
values of $r$ by analytic continuation.  We remark that these analytic
continuations are not necessarily physical states for all values of
the parameters $r,q$.

\subsection{Vertex operators}

In order to diagonalize (\ref{RENOR}) for general values of $q$, with
$-1<q<0$, we follow the strategy proposed in \cite{DFJMN}.
Figure~\ref{f2} suggests (see chapter 4 of \cite{JM}) that we identify
$\H^{(i)}$ with the integrable highest weight module $V(\Lambda_i)$ of
$\uq$, $i=0,1$, and the horizontal lines with the components
$\Phi_\vep^{(1-i,i)}(\z)$ of the vertex operator of type I:
\begin{equation}\label{eqn:VOI}
\Phi^{(1-i,i)}(\z):\H^{(i)}\rightarrow \H^{(1-i)}\otimes V,
\qquad \Phi^{(1-i,i)}(\z)=\sum_\vep \Phi^{(1-i,i)}_\vep(\z)\otimes v_\vep.
\end{equation}
In the present case, the algebra $\uq$ does not play the role of a
symmetry algebra, as it does not commute with the Hamiltonian.
Accordingly we shall treat ${\cal H}^{(i)}$ and
$\Phi^{(1-i,i)}_\vep(\z)$ simply as objects which enjoy the properties
summarized in Appendix A, and disregard their representation
theoretical meaning.

The point of using the vertex operators is that they are well defined
objects, free from the difficulty of divergence.  We define the
`renormalized' transfer matrix
\begin{equation}\label{eqn:TB}
T_B^{(i)}(\z)~= ~
g~ \sum_{\vep,\vep'} \Phi^{*(i,1-i)}_{\vep}(\z^{-1})K_\vep^{\vep'}(\z)
\Phi^{(1-i,i)}_{\vep'}(\z)~,
\end{equation}
(cf. (\ref{eqn:TBfin})).  Here,
\[
\Phi^{*(i,1-i)}_\vep(\z)=\Phi^{(i,1-i)}_{-\vep}\Bigl(-q^{-1}\z\Bigr),
\qquad
g=\frac{\pr{q^2;q^4}}{\pr{q^4;q^4}}.
\]
Henceforth we shall suppress the
label $i$ for the sectors when there is no fear of confusion.
The renormalized Hamiltonian $H_B$ is then defined through the formula
(cf. (\ref{eqn:HviaT}))
\be\lb{DERHAM}
\frac{d}{d\z}T_B(\z)\Bigl|_{\z=1}=\frac{4q}{1-q^2}H_B~.
\en

Using the commutation relations of type I vertex operators
(\ref{eqn:comI}), along with the invertibility (\ref{eqn:invert}),
one can show
the following properties of the transfer matrix (\ref{eqn:TB}):
\begin{eqnarray}
&&[T_B(\z),T_B(\z')]=0\qquad  (\forall \z,\z'),
\label{eqn:comT}\\
&&T_B(1)=\id,\qquad T_B(\z)T_B(\z^{-1})=\id,
\label{eqn:unitT}\\
&&T_B\bigl(-q^{-1}\z^{-1}\bigr)=T_B(\z).
\label{eqn:crossT}
\end{eqnarray}
These relations reduce, respectively, to the boundary YBE
(\ref{eqn:BYBE}), and the boundary unitarity and crossing relations
(\ref{eqn:bunitarity}), (\ref{eqn:bcrossing}) for the $K$-matrix.

\subsection{Energy levels}
Consider now the eigenvalues of the transfer matrix,
\begin{equation}
T_B(\z)\ket{v}=t(\z)\,\ket{v}~,\lb{eig}
\end{equation}
where $\ket{v}$ is some eigenvector of the transfer matrix.
The relations (\ref{eqn:comT})--(\ref{eqn:crossT}) imply
\begin{equation}\label{eqn:relE}
t(1)=1,\quad t(\z)t(\z^{-1})=1,\quad
t\bigl(-q^{-1}\z^{-1}\bigr)=t(\z).
\end{equation}

Multiplying $\Phi_\vep(\z^{-1})$ to \refeq{eig} from the left, and
using the inversion relation (\ref{eqn:invert}) and the definition of
the transfer matrix (\ref{eqn:TB}), we see that
the eigenvalue problem \refeq{eig}
is equivalent to
\begin{equation}\label{eqn:eigVO}
\sum_{\vep'}K^{\vep'}_{\vep}(\z)\Phi_{\vep'}(\z)\ket{v}=
t(\z)~\Phi_\vep(\z^{-1})\ket{v}~,
\qquad  (\vep=\pm)~.
\end{equation}

Consider the region of the parameters where
\be\lb{REGI}
1<\z<1+\vep,\quad-\vep<q<0,\quad-1<r<-1+\vep,
\en
where $\vep>0$ is sufficiently small. Since $\z>1$, the lowest
eigenvalue of the Hamiltonian $H_B$ corresponds to the largest
eigenvalue of the transfer matrix $T_B(\z)$ (see \refeq{DERHAM}, and
note that $q<0$).  In view of the study of $\hzero$ in Section 2.2, we
assume that 1)
the largest eigenvalue of $T^{(i)}_B(\z)$
has multiplicity one, and 2) the corresponding eigenvector has
an expansion at $q=0$, starting from $\ket{p^{(i)}}$.

Let $\Lambda^{(i)}(\z)$ denote the eigenvalue corresponding to the vacuum
vectors $\ket{i}_B$ of Section 2.2. Then
in the region
\refeq{REGI}, $\Lambda^{(0)}(\z)>\Lambda^{(1)}(\z)$,
and $\Lambda^{(0)}(\z)$ is the largest eigenvalue within ${\cal
H}^{(0)}$.  These statements are not necessarily true throughout the
region
\be
1<\z,\quad-1<q<0, \quad-1<r<1.
\en
We infer that
\be\lb{SOL0}
\Lambda^{(0)}(\z)=1.
\en
The reason is as follows. In the region \refeq{REGI}, this is
the largest eigenvalue of the transfer matrix.
We assume $\Lambda^{(0)}(\z)$ is analytic when $1\le |\z^2|\le |q^{-1}|$.
Then the solution to \refeq{relE} is unique, and it is \refeq{SOL0}.

As supporting evidence, we use \refeq{SOL0} to construct
$\ket{0}_B$ in Section 3. We then use the result to calculate, in Appendix
B, the $q$-expansion of $\ket{0}_B$ (for $r=-1$), up to the order
$q^3$. We find that this agrees with the $q$-expansion of the unique
eigenvector of $H_B$ of the form $|p^{(0)}\rangle+O(q)$, as required
in Section 2.3. Therefore we conclude that \refeq{SOL0} is the
correct lowest eigenvalue.

Next we determine $\Lambda^{(1)}(\z)$.
Set
\begin{equation}\label{eqn:ev1}
\Lambda(\z;r)=\frac{K^+_+(\z;r)}{K^-_-(\z;r^{-1})}
=\frac{1}{\z^2}
\frac{\Theta_{q^4}(r\z^2)}{\Theta_{q^4}(r\z^{-2})}
\frac{\Theta_{q^4}(q^2r\z^{-2})}{\Theta_{q^4}(q^2 r\z^{2})}~,
\end{equation}
where $\Theta_p(z)$ is given in (\ref{eqn:tau}).
We now show that
\be\lb{EIG1}
\Lambda^{(1)}(\z;r)=\Lambda(\z;r).
\en

We exploit the spin-reversal symmetry to reduce the calculation of
$\ket{1}_B$ to that of $\ket{0}_B$.  Let
$\nu:\H^{(0)}\rightarrow \H^{(1)}$ be the vector space isomorphism
corresponding to the Dynkin diagram symmetry (see \cite{JM}). Then
\be
\nu~ \Phi^{(0,1)}_\vep(\z)~ \nu~=~ \Phi^{(1,0)}_{-\vep}(\z)~.
\label{eqn:symvertex}
\en
Noting the relation
\[
\sigma^x K(\z;r)\sigma^x~=~\Lambda(\z;r)K(\z;r^{-1})~,
\]
we find that
\begin{equation}\label{eqn:T01}
\nu^{-1}~ T^{(1)}_B(\z;r)~\nu~=~\Lambda(\z;r)T^{(0)}_B(\z;r^{-1})~.
\end{equation}

If $r$ is close to $-1$, the eigenvalue $1$ of $T^{(0)}_B(\z;r^{-1})$ is the
largest. Hence the largest eigenvalue of $T^{(1)}_B(\z;r)$ is
given by \refeq{EIG1}

Now, let us compare the two energy levels $e^{(0)}(r)$ and $e^{(1)}(r)$
of $H_B$, corresponding to $\ket{0}_B$ and $\ket{1}_B$, respectively.
Parameterize $r$ as
\[
r=\cases{-(q^2)^\alpha&if $-1<r<0$;\cr(q^2)^\alpha&if $0<r<1$,\cr}
\]
where $\alpha$ is real and positive.
{}From \refeq{SOL0} and  \refeq{DERHAM}, we have clearly
\[
e^{(0)}(r)=0.
\]
Hence the energy difference is simply
\[
\Delta e(r)=e^{(1)}(r)-e^{(0)}(r)=e^{(1)}(r).
\]
{}From \refeq{ev1} and (\ref{eqn:DERHAM}), we find
\begin{equation}\label{eqn:e1}
\Delta e(r)=\cases{\epsilon(1){\rm sn}(2K'\alpha,k')
&if $-1\le r<0$;\cr
{\displaystyle{\epsilon(1)\over k'{\rm sn}(2K'\alpha,k')}}
&if $0<r<1$,\cr}
\end{equation}
where the Jacobi elliptic functions refer to the nome
$-q=e^{-\pi K'/K}$ (see 7.5 in \cite{JM}), and
\begin{eqnarray}
&&\epsilon(\xi)={2K\over\pi}{\rm sinh\,}{\pi K'\over K}{\rm dn}
\left({2K\over\pi}\theta,k\right),
\qquad \xi=-ie^{i\theta},
\label{eqn:eps}\\
&&\epsilon(1)={2Kk'\over\pi}{\rm sinh\,}{\pi K'\over K}.
\label{eqn:epsmin}
\end{eqnarray}

\begin{figure}[t]
\epsfxsize=4in
\centerline{\epsffile{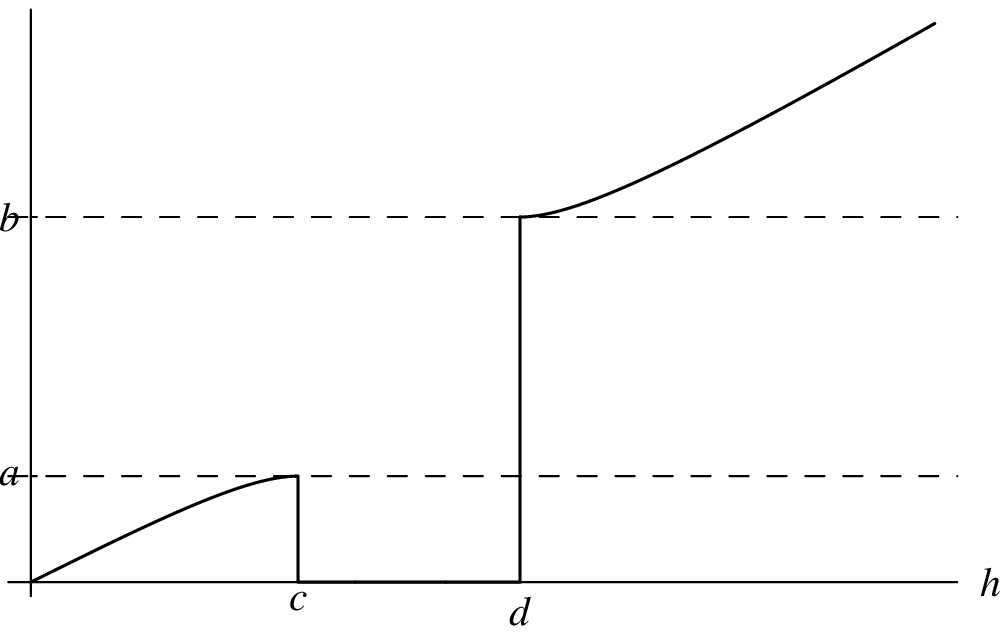}}
\caption{The energy difference $\Delta e(r)$
between the vacuum states as a function of
$h$, when $q=-0.15$.
The energy $a=\epsilon(1)$, and $b=\epsilon(1)/k'$. The critical field
$c=\hca$ and $d=\hcb$.}
\label{f3}
\end{figure}

In Figure~\ref{f3}, the energy difference $\Delta e(r)$ is plotted
as a function of $h$. The energy increases monotonically from $0$ to
$\epsilon(1)$ (which is the mass gap, see below) when $0\le h<\hca$, and from
$\epsilon(1)/k'$ to $\infty$ when $\hcb<h\le\infty$. In the latter
region, the energy of $\ket{1}_B$ is greater than the mass gap.  The
possible existence of such a situation, where the boundary bound state
energy is greater than the single particle energy, was mentioned
in~\cite{GZ}, although it does not occur in the Ising model. When
$\hca<h<\hcb$, the state $|1\rangle_B$ is well defined, as we shall
argue in Section 3.1.

In the next section we shall find the vacuum vectors $\ket{0}_B$ and
$\ket{1}_B$ explicitly by using the bosonization formulas for the
vertex operators. Once the vacuum vectors are found, it is possible to
create the excited states by application of the vertex operators of
type II, $\Psi_{\mu}^*(\xi)$ ($\mu=\pm$), in much the same way as in
the bulk theory (see \cite{JM}).  The commutation relations
(\ref{eqn:phipsi}) imply that
\[
\Psi_{\mu_1}^*(\xi_1)\cdots \Psi_{\mu_m}^*(\xi_m)\ket{i}_B
\]
is an eigenstate of $T_B(\z)$ with eigenvalue
$\Lambda^{(i)}(\z)\times \prod_{j=1}^m \tau_B(\z,\xi_j)$, where
\be\lb{BONEP}
\tau_B(\z,\xi)=\tau(\z/\xi)\tau(\z\xi)
\en
and $\tau(\z)$ is given in (\ref{eqn:tau}).
The single particle
energy is therefore given by $e^{(i)}(r)+\epsilon(\xi)$, with its
minimum value being $e^{(i)}(r)+\epsilon(1)$. Therefore the mass gap
is $\e(1)$.

\setcounter{equation}{0}
\section{Vacuum vectors and the boundary $S$-matrix}
\subsection{Vacuum vectors}

The perturbative calculation of Appendix B suggests that the vacuum
vector $\ket{i}_B$ is uniquely determined by the relation
\be
T^{(i)}_B(\z)\ket{i}_B=\Lambda^{(i)}(\z)\ket{i}_B.
\en
In this section we invoke the bosonization method to find the
explicit formulas for them, assuming uniqueness.

First consider the case $\ket{0}_B$.
Since the total spin is conserved,
it should be a (possibly infinite) linear combination of the states created
by the oscillators $a_{-n}$ ($n>0$) over the highest weight vector
$\ket{0}$.
We make the ansatz that it has the form (\ref{eqn:grst}), with
\[
\ket{0}_B=e^{F_0}\ket{0},
\qquad
F_0=\frac{1}{2}\sum_{n=1}^\infty \frac{n\alpha_n}{[2n][n]} a_{-n}^2
+\sum_{n=1}^\infty \beta_n^{(0)} a_{-n}.
\]
Note that the coefficients $\alpha_n, \beta_n^{(0)}$ cannot depend on
$\z$ due to the commutativity of the transfer matrices.
Using the bosonization formula (\ref{eqn:phim}), equation
(\ref{eqn:eigVO}), with $t(\z)=\Lambda^{(0)}(\z)=1$ and $\vep=-$,
becomes
\[
\varphi(z;r) e^{P(z)}e^{Q(z)}e^{F_0}\ket{0}=(~z\leftrightarrow z^{-1}~).
\]
The presence of $e^{F_0}$ has the effect of a Bogoliubov transformation
\begin{equation}\label{eqn:Bogo}
e^{-F_0}a_ne^{F_0}=a_n+\alpha_n a_{-n}+\frac{[2n][n]}{n}\beta_n^{(0)},
\qquad
e^{-F_0}a_{-n}e^{F_0}=a_{-n}.
\end{equation}
By a straightforward calculation, the coefficients $\alpha_n$ and
$\beta_n^{(0)}$ are found to be
\[
\alpha_n=-q^{6n},
\qquad
\beta_n^{(0)}= -\frac{q^{7n/2}r^n}{[2n]}
-\theta_n\frac{q^{5n/2}(1-q^{n})}{[2n]}~,
\]
where
\[
\theta_n=\cases{1&if $n$ is even;\cr 0 &if $n$ is odd.\cr}
\]

We must verify that the state $\ket{0}_B$ found in this manner satisfies
also the relation (\ref{eqn:eigVO}) with $\vep=+$. Using
\begin{eqnarray*}
e^{Q(z)}\ket{0}_B&=&\varphi(z^{-1};r)e^{P(z^{-1})}\ket{0}_B, \\
e^{S^-(w)}\ket{0}_B&=&
(1-q^6w^{-2})(1-q^4rw^{-1})e^{R^-(q^6w^{-1})}\ket{0}_B~,
\end{eqnarray*}
the verification is reduced to showing that
\[
\oint\frac{dw}{2\pi i w}\frac{(1-rz)(w^2-q^6)(w-rq^4)}
{(w-q^2z)(w-q^4z)(w-q^4z^{-1})} e^{R^-(w)+R^-(q^6w^{-1})}\ket{0}
=(~z\leftrightarrow z^{-1}~).
\]
Here the contour encircles $w=0, q^4z^{\pm 1}$ but not $q^2z^{\pm 1}$.
The desired equality can be shown by the change of variable
$w\rightarrow q^6w^{-1}$.

Similar calculations give the other vacuum vector $\ket{1}_B$, as well
as the vacuum vectors in the dual space, ${\cal H}^{*(i)}$,
defined by
\[
{}_B\bra{i}T_B(\z)=\Lambda^{(i)}(\z){}_B\bra{i}~.
\]
The analog of (\ref{eqn:eigVO}) is
\begin{equation}\label{eqn:leigVO}
\sum_{\vep}{}_B\bra{i}\Phi^{*}_\vep(\z^{-1})K^{\vep'}_\vep(\z)
=\Lambda^{(i)}(\z)~{}_B\bra{i}\Phi^*_{\vep'}(\z),
\qquad (\vep'=\pm)~.
\end{equation}
The results are summarized as follows:
\begin{eqnarray}
&&\ket{i}_B=e^{F_i}\ket{i},
\qquad
F_i=\frac{1}{2}\sum_{n=1}^\infty \frac{n\alpha_n}{[2n][n]} a_{-n}^2
+\sum_{n=1}^\infty \beta_n^{(i)} a_{-n},
\label{eqn:Fi}\\
&&{}_B\bra{i}=\bra{i}e^{G_i},
\qquad
G_i=\frac{1}{2}\sum_{n=1}^\infty \frac{n\gamma_n}{[2n][n]} a_{n}^2
+\sum_{n=1}^\infty \delta_n^{(i)} a_{n},
\label{eqn:Gi}
\end{eqnarray}
where
\begin{eqnarray}
&&\alpha_n=-q^{6n},\qquad \gamma_n=-q^{-2n},
\label{eqn:alphan}\\
&&
\beta_n^{(i)}=-\theta_n\frac{q^{5n/2}(1-q^{n})}{[2n]}
+\cases{ -\displ{\frac{q^{7n/2}r^n}{[2n]}} & ($i=0$);\cr
         +\displ{\frac{q^{3n/2}r^{-n}}{[2n]}} & ($i=1$),\cr}
\label{eqn:betan}\\
&&
\delta_n^{(i)}=\theta_n\frac{q^{-3n/2}(1-q^{n})}{[2n]}
+\cases{ -\displ{\frac{q^{-5n/2}r^n}{[2n]}} & ($i=0$);\cr
         +\displ{\frac{q^{-n/2}r^{-n}}{[2n]}} & ($i=1$).\cr}
\label{eqn:deltan}
\end{eqnarray}

The relation (\ref{eqn:T01}) suggests that the two vacuum vectors
should be related by
\be
\nu\left(\ket{0}_B\right)\Bigr|_{r\rightarrow r^{-1}}=\ket{1}_B.
\label{eqn:symvac}
\en
However, since the spin-reversal symmetry is obscured in the bosonization,
we do not know how to verify this directly from the explicit formulas
(\ref{eqn:Fi}).

\medskip\noindent
{\sl Remark.}\quad
It is useful to generalize (\ref{eqn:TB}) slightly and consider
\begin{equation}\label{eqn:TBt}
T_B^{(i)}(\z;t)=
g \sum_{\vep,\vep'} \Phi^{*(i,1-i)}_{\vep}(t\z^{-1})K_\vep^{\vep'}(\z)
\Phi^{(1-i,i)}_{\vep'}(t\z),
\end{equation}
which corresponds to changing the spectral parameters for the vertical
lines from $1$ to $t^{-1}$.
The properties (\ref{eqn:comT})--(\ref{eqn:crossT}) are
unaffected by this change.
The vacuum vectors for (\ref{eqn:TBt}) are obtained by replacing
\[
a_n\rightarrow t^{-2n}a_n,
\quad
a_{-n}\rightarrow t^{2n}a_{-n}
\qquad (n>0)
\]
in (\ref{eqn:Fi}), (\ref{eqn:Gi}).
Doing so and specializing $r=\z^{-2}$ ($i=0$) or $r=\z^2$ ($i=1$)
we can drop one of the terms in (\ref{eqn:TBt}), to obtain
\begin{eqnarray*}
&&
{}_B\bra{0}\Phi^{*}_-(t\z^{-1})\Phi_-(t\z)=\lambda(\z^2){}_B\bra{0},
\quad
\Phi^{*}_-(t\z^{-1})\Phi_-(t\z)\ket{0}_B=\lambda(\z^2)\ket{0}_B,
\\
&&
{}_B\bra{1}\Phi^{*}_+(t\z^{-1})\Phi_+(t\z)=\lambda(\z^2){}_B\bra{1},
\quad
\Phi^{*}_+(t\z^{-1})\Phi_+(t\z)\ket{1}_B=\lambda(\z^2)\ket{1}_B,
\end{eqnarray*}
with
\[
\lambda(z)=\frac{(q^8z^2;q^8)_\infty}{(q^6z^2;q^8)_\infty}
\frac{(q^4z^{-2};q^8)_\infty}{(q^2z^{-2};q^8)_\infty}.
\]
These formulas play a role in determining the asymptotic behavior
of the iteration of the vertex operators, conjectured in \cite{DFJMN}
and proved recently in \cite{Etingof}.
\medskip

Next, consider the small $q$ expansion of the vectors
$\ket{i}_B~(i=0,1)$. As stated in Section 1, we examine whether they
exist or not in the sense they are regular at $q=0$.  Here, we
say a vector is regular if it belongs to the upper crystal lattice of
Kashiwara (see (\ref{eqn:APP1})).

The expression \refeq{Fi} for $\ket{i}_B$ is not suitable for this
purpose, since we do not know a simple criterion as to whether a
vector expressed in terms of bosons belongs to the crystal lattice. In
Appendix B, we rewrite $\ket{0}_B$ in terms of the global base
vectors. Although we have no proof to all orders, the expansion
suggests that $\ket{0}_B$ is well defined when $h\ge 0$, but not
when $|r|\ge1/|q|$, i.e., $-\hcb\le h\le-\hca$.  The expansion of
$\ket{1}_B$ is obtained by the replacement of $r$ to $r^{-1}$ in
\refeq{APP1}. Thus, we observe three regions for $h\ge0$:
\par\noindent
(i)\quad $0\le h<\hca$: Both $\ket{0}_B$ and $\ket{1}_B$
are well defined;
\par\noindent
(ii)\quad $\hca\le h\le\hcb$: $\ket{0}_B$ is
well defined, but $\ket{1}_B$ is not well defined;
\par\noindent
(iii)\quad $\hcb<h\le\infty$: Both $\ket{0}_B$ and $\ket{1}_B$
are again well defined.
\par\noindent

\subsection{Boundary $S$-matrix}
The two particle $S$-matrix is unchanged from the bulk theory:
\begin{eqnarray*}
&&\Psi_{\mu_1}^*(\xi_1)\Psi_{\mu_2}^*(\xi_2)\ket{i}_B
=\sum_{\mu_1',\mu_2'}S^{\mu_1'\,\mu_2'}_{\mu_1\,\mu_2}(\xi_1/\xi_2)
\Psi_{\mu_2'}^*(\xi_2)\Psi_{\mu_1'}^*(\xi_1)\ket{i}_B,
\\
&&
S(\xi)=-R(\xi)^{t_1t_2}=-R(\xi).
\end{eqnarray*}

The formula \refeq{BONEP} shows that the eigenvalues corresponding to
the states $\Psi^*_\mu(\xi)\ket{i}_B$ and
$\Psi^*_\mu(\xi^{-1})\ket{i}_B$ are equal. In fact, these states are
proportional to each other.  Using the bosonic expressions for the
states ${}_B\bra{i}$ and $\ket{i}_B$, we find
\begin{eqnarray}
&&\Psi^*_\mu(\xi)\ket{i}_B
=
M^{(i)\mu}_\mu(\xi)\Psi^*_\mu(\xi^{-1})\ket{i}_B,
\label{eqn:M}\\
&&{}_B\bra{i}\Psi_\mu(\xi^{-1})
=
{}_B\bra{i}\Psi_\mu(\xi)M^{(i)\mu}_\mu(\xi).
\label{eqn:dM}
\end{eqnarray}
We therefore interpret $M^{(i)}(\xi)$ as the boundary $S$-matrix.  It
is computed as in the previous subsection, where we now use
(\ref{eqn:psim}) and (\ref{eqn:psip}).  We find
\begin{equation}\label{eqn:Mi}
M^{(0)\mu}_\mu(\xi)=M^{\mu}_\mu(\xi;r),
\qquad
M^{(1)\mu}_\mu(\xi)=M^{-\mu}_{-\mu}(\xi;r^{-1}),
\end{equation}
where
\begin{equation}\label{eqn:Mr}
M(\xi;r)=\frac{1}{\overline{f}(\xi;r)}
\left(\matrix{
\displaystyle\frac{1-rq^{-1}\xi^2}{\xi^2-rq^{-1}} & 0 \cr
0 & 1 \cr}
\right)
\end{equation}
and
\[
\overline{f}(\xi;r)=
-\xi^2
\frac{\overline{\varphi}(\xi^{-2};r)}{\overline{\varphi}(\xi^{2};r)},
\qquad
{\overline{\varphi}(z;r)}
=
\frac{\pr{q^3rz;q^4}}{\pr{qrz;q^4}}
\frac{\pr{q^2z^2;q^8}}{\pr{q^8z^2;q^8}}.
\]
Notice that, up to a scalar factor, the $M$-matrix is of the same form
as the $K$-matrix (\ref{eqn:K}), except for the shift $r\rightarrow
rq^{-1}$. The $M$-matrix has properties similar to those of the
$K$-matrix:
\begin{eqnarray}
&&M_2(\xi_2)S_{12}(\xi_2\xi_1)M_1(\xi_1)S_{21}(\xi_1/\xi_2)
=
S_{12}(\xi_1/\xi_2)M_1(\xi_1)S_{21}(\xi_1\xi_2)M_2(\xi_2),\nonumber
\\
&&M(\xi)M(\xi^{-1})=1,\nonumber
\\
&&M_a^b(-q^{-1}\xi^{-1})=
\sum_{a',b'}S^{b'\,-a'}_{-b\,a}(-q\xi^2)M^{a'}_{b'}(\xi).\label{smatrix}
\end{eqnarray}
These formulas can be obtained either directly, or by using
(\ref{eqn:M}), (\ref{eqn:dM}). For instance, the last relation follows
if we multiply (\ref{eqn:M}) by $\Psi_{\nu}(\xi')$ from the left and
take the residue at $\xi'=\xi$ using the property (\ref{eqn:delta}).

In \cite{GZ}, similar equations to (\ref{smatrix}) were proposed, as
the bootstrap conditions for boundary field theories. Here, these
equations are derived, in the context of the lattice model.

We now consider the analyticity properties of the boundary $S$-matrix,
whose singularity structure determines the existence of the boundary
bound states.

First, recall the discussion of the Ising model (in the continuum
limit) in~\cite{GZ}.  The boundary $S$-matrix is given by
\be
R_h(\theta)=i\tanh\Bigl({\pi i\over4}-{\theta\over2}\Bigr)
\frac{\kappa-i\sinh\theta}{\kappa+i\sinh\theta}
\en
where $\theta$ is the rapidity of the particle with mass $m$, and
$\kappa=1-h^2/2m$, $h$ being the boundary magnetic field.  The result
is as follows: If $0\le h<\sqrt{2m}$, the pole of $R_h(\theta)$ at
$\theta={\pi i\over2}-iv$, where $\kappa=\cos v$, stays in the
physical strip, $0<{\rm Im}~\theta\le\pi$.  When $h\ge\sqrt{2m}$, the
pole moves away from the physical strip.

In our case, the pole at $\xi^2=rq^{-1}$ is in the physical strip,
$1<|\xi^2|\le|q|^{-2}$, when $-1\le r<q$ ($0\le h < \hca$), but also
when $-q<r\le1$ ($\hcb<h\le\infty$). Here, we considered only the
region $-1\le r\le1$ ($h\ge0$).  These two regions are the regions (i)
and (iii) discussed in Section 3.1.

We recognize that $\ket{1}_B$ is the bound state corresponding to the
above mentioned pole, by using the bosonic expressions for the vacuum
and single--particle vectors.  The single--particle state
$\Psi^*_+(\xi)\ket{0}_B$ has a series of simple poles at
\[
\xi^2=q^{-1}r;~~q^{-1}r^{-1},~q^{-5}r^{-1},~\cdots~;~~
q^3r,~q^7r,~\cdots~.
\]
By explicit computation, we find that the state $\ket{1}_B$ can be
obtained as the residue at the first pole:
\begin{eqnarray}
&&{\rm Res}_{\xi^2=q^{-1}r}
\left({}_B\bra{0}\Psi_+(\xi^{-1})\frac{d\xi^2}{\xi^2}\right)
=g~c(r^{-1})~{}_B\bra{1},
\label{eqn:bdst1}\\
&&{\rm Res}_{\xi^2=q^{-1}r}
\left(\Psi_+^*(\xi)\ket{0}_B\frac{d\xi^2}{\xi^2}\right)
=g~c(r^{-1})~\ket{1}_B,
\label{eqn:bdst2}
\end{eqnarray}
where
$c(z)=(q^2z^2;q^8)_\infty/(q^4z^2;q^8)_\infty$.
The equations (\ref{eqn:bdst1}), (\ref{eqn:bdst2})
are also valid if we replace
${}_B\bra{0}\Psi_+(\xi^{-1})\rightarrow {}_B\bra{1}\Psi_-(\xi^{-1})$,
$\Psi_+^*(\xi)\ket{0}_B\rightarrow \Psi_-^*(\xi)\ket{1}_B$
and $r\rightarrow r^{-1}$.
The formulas above allow us to interpret $\ket{i}_B$
as the `boundary bound states'
in the regions (i), (iii) mentioned above.
\setcounter{equation}{0}
\section{Correlation functions}

\subsection{$N$-point function}
In this section we calculate the vacuum expectation values of products
of type I vertex operators, and obtain them as integrals of
meromorphic functions involving infinite products.  Upon
specialization of the spectral parameters, they give multi-point
correlation functions of the local spin operators of the $\XXZ$ chain
with a boundary interaction.

We will consider the following $N$-point function with $N$ even:
\begin{equation}\label{eqn:P}
P^{(i)}_{\vep_1\ldots\vep_N}(\z_1,\ldots,\z_N)=
\frac{
{}_B\bra{i}
\Phi_{\varepsilon_1}^{(i,1-i)}(\zeta_1)\Phi_{\varepsilon_2}^{(1-i,i)}(\zeta_2)
\cdots \Phi_{\varepsilon_N}^{(1-i,i)}(\zeta_N)\ket{i}_B}
{{}_B\langle i|i \rangle_B}.
\end{equation}
Fixing $\{\vep_1,\ldots,\vep_N\}$, let us denote by $A$ the index set
\[
A=\{ j \mid 1\leq j\leq N, \vep_j=+1\}.
\]
Since the total spin is conserved, (\ref{eqn:P}) is non-trivial only if
$\sum_{j=1}^N\vep_j=0$, in which case $A$ has $N/2$ elements.

We remark that, as a consequence of the relation of the two vacuum
vectors (\ref{eqn:symvac}) and the one for the vertex operators
(\ref{eqn:symvertex}), one obtains
\be
P^{(1)}_{\vep_1\ldots\vep_N}(\z_1,\ldots,\z_N)=
P^{(0)}_{-\vep_1\ldots-\vep_N}(\z_1,\ldots,\z_N)\Bigr|_{r\to r^{-1}}.
\label{eqn:Symm}
\en

In order to evaluate the expectation values (\ref{eqn:P}) we invoke
the bosonization formulas (\ref{eqn:phim}), (\ref{eqn:phip}),
(\ref{eqn:PQ}), (\ref{eqn:RS}).
By normal-ordering the product of vertex operators, we have
\begin{eqnarray}
\lefteqn{
P^{(i)}_{\vep_1\ldots\vep_N}(\z_1,\ldots,\z_N)
}\cr
&&=(-q^3)^{N^2/4+iN/2-\sum_{a\in A}a}
(1-q^2)^{N/2 }
\displaystyle{\prod_{j=1}^{N}{\zeta_j}^{\frac{1+\epsilon_j}{2}-j+N+i}}
\displaystyle{\prod_{j<k}}
\frac{\displaystyle{(q^2z_k/ z_j;q^4)_{\infty}}}
     {\displaystyle{(q^4z_k/ z_j;q^4)_{\infty}}}
\cr
&&\qquad\times\displaystyle{\prod_{a\in A}\oint_{C_{a}}
\frac{dw_a w_a^{1-i}}{2 \pi \sqrt{-1}}}\
\frac{\displaystyle \prod_{a<b}(w_a-w_b)(w_a-q^2w_b)}
    {\displaystyle \prod_{j \leq a}(z_j-q^{-2}w_a)
                    \prod_{a \leq j}(w_a-q^4 z_j)}\cr
&&\qquad \times I^{(i)}(\{z_j\},\{w_a\}),
\label{eqn:Normp}
\end{eqnarray}
where each  contour $C_a$ is the same one
defined in (\ref{eqn:phip}) as $C_1$.
Here,  $z_j=\z_j^2$ and
\begin{eqnarray*}
&&{I^{(i)}(\{z_j\},\{w_a\})}
=
\frac{_B\bra{i}\exp\left(\sum_{n=1}^{\infty}a_{-n}X_n\right)
\exp\left(-\sum_{n=1}^{\infty}a_{n}Y_n\right)\ket{i}_B}
{{}_B\langle i|i \rangle_B},
\\
&&X_n=\frac{q^{7n/2}}{[2n]}\sum_{j=1}^Nz_j^n-\frac{q^{n/2}}{[n]}\sum_{a}w_a^n,
\qquad
Y_n=\frac{q^{-5n/2}}{[2n]}\sum_{j=1}^Nz_j^{-n}-\frac{q^{n/2}}{[n]}\sum_{a}
w_a^{-n}.
\end{eqnarray*}

Next let us calculate the expectation value $I^{(i)}(\{z_j\},\{w_a\})$.
By using the Bogoliubov transformation (\ref{eqn:Bogo}), we obtain
\begin{eqnarray*}
\lefteqn{I^{(i)}(\{z_j\},\{w_a\})}\\
&&=
\frac{\bra{i}e^{G_i}e^{F_i}
\exp\Bigl(\sum_{n=1}^{\infty}a_{-n}(X_n-\alpha_nY_n)\Bigr)\ket{i}}
{_B\langle i|i \rangle_B}
\\
&&\times
\exp\Bigl(-\sum_{n=1}^{\infty}\frac{[2n][n]}{n}\beta_n^{(i)}Y_n\Bigr)
\exp\Bigl(\frac{1}{2}\sum_{n=1}^{\infty}\frac{[2n][n]}{n}\alpha_nY_n^2\Bigr).
\end{eqnarray*}
We then insert between $e^{G_i}$ and $e^{F_i}$
the completeness relation of the coherent states (\ref{eqn:comp})
and use the integration formula (\ref{eqn:Cohint}). As a result we find
\begin{eqnarray}
\lefteqn{
I^{(i)}(\{z_j\},\{w_a\})
\times\prod_{n=1}^{\infty}(1-\alpha_n\gamma_n)^{1/2}
}\cr
&=&
\exp\Biggl(\sum_{n=1}^{\infty}\frac{[2n][n]}{n}\frac{1}{1-\alpha_n\gamma_n}
\Bigl\{\frac{1}{2}\gamma_nX_n^2-\alpha_n\gamma_nX_nY_n\cr
&&
+\frac{1}{2}\alpha_nY_n^2 +(\delta_n^{(i)}+\gamma_n\beta_n^{(i)})X_n
-(\beta_n^{(i)}+\alpha_n\delta_n^{(i)})Y_n
\Bigr\}\Biggr),
\label{eqn:exI}\\
\lefteqn{
_B\langle i|i \rangle_B
\times\prod_{n=1}^{\infty}(1-\alpha_n\gamma_n)^{1/2}
}\cr
&=&
\exp\Biggl(\frac{1}{2}\sum_{n=1}^{\infty}\frac{[2n][n]}{n}\frac{1}{1-\alpha_n\gamma_n}
\Bigl\{\gamma_n\beta_n^{(i)2}+
\alpha_n\delta_n^{(i)2} +2\beta_n^{(i)}\delta_n^{(i)}
\Bigr\}\Biggr).
\label{eqn:Norm}
\end{eqnarray}
The sums in the right hand sides
are evaluated by making use of the following formulas.
\begin{eqnarray}
\exp\Bigl(
\sum_{n=1}^{\infty}\frac{[n]}{n}q^{-5n/2}z^n\beta^{(i)}_n\Bigr)
&=&\left(\frac{\prq{q^4z^2}}{\prq{q^2z^2}}\right)^{-1/2}
\left\{
 \begin{array}{ll}
          \varphi(z;r)^{-1},& \\
          {\Bigl[(1-r^{-1}z) \varphi(z;r^{-1})}\Bigr]^{-1},&\\
 \end{array}
\right. \cr
\exp\Bigl(
\sum_{n=1}^{\infty}\frac{[2n]}{n}q^{an}w^n\beta^{(i)}_n\Bigr)
&=&\left(\frac{1-q^{2a+5}w^2}{1-q^{2a+7}w^2}
\right)^{1/2}
\left\{
 \begin{array}{ll}
          1-q^{a+7/2}rw,& \\
          (1-q^{a+3/2}r^{-1}w)^{-1},
          & \\
          \end{array}
\right.\cr
\exp\Bigl(
\sum_{n=1}^{\infty}\frac{[n]}{n}q^{3n/2}z^n\delta^{(i)}_n\Bigr)
&=&\left(\frac{\prq{q^4z^2}}{\prq{q^2z^2}}
\right)^{1/2}
\left\{
 \begin{array}{ll}
          {(1-rz)\varphi(z;r)},& \\
          \varphi(z;r^{-1}),
          & \\
\end{array}\right. \cr
\exp\Bigl(
\sum_{n=1}^{\infty}\frac{[2n]}{n}q^{an}w^n\delta^{(i)}_n\Bigr)
&=&\left(\frac{1-q^{2a-1}w^2}{1-q^{2a-3}w^2}
\right)^{1/2}
\left\{
 \begin{array}{ll}
          1-q^{a-5/2}rw,& \\
          (1-q^{a-1/2}r^{-1}w)^{-1}.
          & \\
\end{array}\right. \nonumber
\end{eqnarray}
In the above, the upper line applies to the case
$i=0$ and the lower to $i=1$.
Let us summarize below the results of computation.

\medskip\noindent
{\it The norm of the vacuum vectors:}
\begin{eqnarray}
{}_B\!\langle 0|0\rangle_B&=&\frac{({q^4r^2};q^8)_\infty}{({q^6};q^8)_\infty
({q^2r^2};q^8)_\infty},\\
{}_B\!\langle 1|1\rangle_B&=&\frac{({q^4/r^2};q^8)_\infty}{({q^6};q^8)_\infty
({q^2/r^2};q^8)_\infty}.
\end{eqnarray}

\medskip
\noindent{\it The N-point function:}
\begin{eqnarray}
\lefteqn{
P^{(i)}_{\vep_1\ldots\vep_N}(\z_1,\ldots,\z_N)
}\cr
&&=(-q^3)^{N^2/4-\sum_{a\in A}a}
  \left(\frac{\{q^6\}_{\infty}}{\{q^8\}_{\infty}}\right)^N
  (q^2;q^2)_{\infty}^{N/2}
  \prod_{j=1}^{N}\zeta_j^{\frac{1+\varepsilon_j}{2}-j+N} \cr
&&\times \displaystyle{\prod_{j<k}}
\frac{\displaystyle{\{q^6z_j z_k\}_{\infty} \{q^6z_j/z_k\}_{\infty}
                    \{q^6z_k/z_j\}_{\infty} \{q^2/z_j z_k\}_{\infty}}}
     {\displaystyle{\{q^8z_j z_k\}_{\infty} \{q^8z_j/z_k\}_{\infty}
                    \{q^8z_k/z_j\}_{\infty} \{q^4/z_j z_k\}_{\infty}}}\cr
&&\times\displaystyle{\prod_{j=1}^N}
\frac{\displaystyle{[q^{10}z_j^2]_{\infty} [q^{14}z_j^2]_{\infty}
                    [q^{10}/z_j^{2}]_{\infty} [q^{6}/z_j^{2}]_{\infty}}}
     {\displaystyle{[q^{12}z_j^2]_{\infty} [q^{16}z_j^2]_{\infty}
                    [q^{12}/z_j^{2}]_{\infty} [q^8/z_j^{2}]_{\infty}}}
\cr
&&\times \displaystyle{
      \prod_{a\in A}\oint_{C_a^{(i)}}\frac{dw_a}{2\pi \sqrt{-1}}}
      \frac{w_a^{N-2a}\displaystyle{
      (q^{-2}w_a^2;q^4)_{\infty}(q^6/w_a^{2};q^4)_{\infty}}}
      {\displaystyle{\prod_{j\leq a}( z_j-q^{-2}w_a)
                    \prod_{a\leq j}(w_a-q^4z_j)}} \cr
&&\times
\frac{\displaystyle{\prod_{a<b}
(q^{-2}w_a w_b;q^2)_{\infty}(q^4w_a/w_b;q^2)_{\infty}
(w_b/w_a;q^2)_{\infty}(q^6/w_a w_b;q^2)_{\infty}}}
{\displaystyle{\prod_a \prod_j
   {(q^2z_jw_a;q^4)_{\infty}(q^8z_j/w_a;q^4)_{\infty}
                 (q^2w_a/z_j;q^4)_{\infty}(q^4/z_jw_a;q^4)_{\infty}}}}\cr
\cr
&&
 \times \left\{
 \begin{array}{ll}
 \displaystyle{\prod_{j=1}^{N}\frac{(q^2 r z_j;q^4)_{\infty}}
                                 {(q^4 r z_j;q^4)_{\infty}}}
              \prod_a \frac{w_a}{1-q^{-2}rw_a},
        &\mbox{(for $i=0$)} \\
(-q^3)^{N/2}\displaystyle{
              \prod_{j=1}^{N}\z_j\frac{(r^{-1} z_j^{-1};q^4)_{\infty}}
                                 {(q^2 r^{-1} z_j^{-1};q^4)_{\infty}}}
              \prod_a \frac{1}{1-q^2r^{-1}w_a^{-1}}.
         & \mbox{(for $i=1$)}
\end{array}\right.
\nonumber\\
&&\label{eqn:Npt}
\end{eqnarray}
Here, the contour $C^{(0)}_a$ encircles the points $q^{4l}z^{-1}_j$
(for all $j$), $q^{4l}z_j\ (a\leq j)$ and $q^{4+4l}z_j\ (a> j),\
l=1,2,3,..$ but not the point $q^2r^{-1}$, whereas the contour
$C^{(1)}_a$ encircles the point $q^2r^{-1}$ in addition to the same
points as $C^{(0)}_a$ does.  We have also set
\[
\{z\}_\infty=(z;q^4,q^4)_\infty,
\qquad
[z]_\infty=(z;q^8,q^8)_\infty
\]
with $(z;p,q)_\infty=\prod_{j,k\ge 0}(1-zp^jq^k)$.

\subsection{Boundary magnetization}
We now specialize the formula (\ref{eqn:Npt}) to obtain the
correlation functions of local operators. In particular,
we obtain the boundary magnetization $\langle \sigma^z_1\rangle$.

Let $L$ be a linear operator on the $n$-fold tensor
product of the two dimensional space $V\otimes \cdots\otimes V$.
The corresponding local operator ${\cal L}$ acting on
our space of states ${\cal H}$ can be defined in terms of the
type I vertex operators,
in exactly the same way as in the bulk theory \cite{JMMN,JM}.
Explicitly, if $L$ is the spin operator at the $n$-th site
\[
\sigma^{\alpha}_n=\sigma^\alpha \otimes
\underbrace{{\rm id}\otimes\cdots\otimes{\rm id}}_{n-1}
\qquad (\alpha=\pm, z),
\]
the corresponding local operator ${\cal L}^{(i)\alpha}$ is given by
\begin{eqnarray}
{\cal L}^{(i)\pm}&=&E^{(i)}_{\pm\mp}(1,\ldots,1),
\label{eqn:spm}\\
{\cal L}^{(i)z}&=&E^{(i)}_{++}(1,\ldots,1)-
E^{(i)}_{--}(1,\ldots,1),
\label{eqn:sz}
\end{eqnarray}
where
\begin{eqnarray}\label{eqn:Eij}
\lefteqn{
E^{(i)}_{\vep\vep'}(\z_1,\ldots,\z_n)
}\cr
&&=g^n\sum_{\vep_1,\ldots,\vep_{n-1}}
\Phi^{*(i,1-i)}_{\vep_1}(\z_1)
\cdots
\Phi^{*(\iota, 1-\iota)}_{\vep_{n-1}}(\z_{n-1})
\Phi^{*(1-\iota, \iota)}_{\vep}(\z_{n})\cr
&&\qquad \times
\Phi^{(\iota, 1-\iota)}_{\vep'}(\z_{n})
\Phi^{(1-\iota, \iota)}_{\vep_{n-1}}(\z_{n-1})
\cdots
\Phi^{(1-i,i)}_{\vep_1}(\z_1)
\end{eqnarray}
with $\iota=i$ for even $n$ and $\iota=1-i$ for odd $n$.

Thanks to the inversion property (\ref{eqn:invert}), the
$n$-point function of the product of successively ordered
operators of the form (\ref{eqn:Eij})
becomes particularly simple:
\begin{eqnarray*}
\lefteqn{_B\bra{i}E^{(i)}_{\vep_n\vep_n'}(\z_1,\ldots,\z_n)
E^{(i)}_{\vep_{n-1}\vep_{n-1}'}(\z_1,\ldots,\z_{n-1})\cdots
E^{(i)}_{\vep_1\vep_1'}(\z_1)\ket{i}_B
}\cr
&&=g^n\,{}_B\bra{i}
\Phi^{*(i,1-i)}_{\vep_1}(\z_1)\cdots
\Phi^{*(1-\iota, \iota)}_{\vep_n}(\z_{n})
\Phi^{(\iota, 1-\iota)}_{\vep_n'}(\z_{n})\cdots
\Phi^{(1-i,i)}_{\vep_1'}(\z_1)
\ket{i}_B.
\end{eqnarray*}
The integral formula for this function is obtained from
(\ref{eqn:Npt}) by the specialization
\begin{eqnarray*}
&&\vep_1,\vep_2,\ldots.,\vep_N\to
-\vep_1,\ldots.,-\vep_n,
\vep_n',\ldots.,\vep_1'\cr
&&\z_1,\z_2,\ldots.,\z_N\to -q^{-1}\z_1,\ldots,-q^{-1}\z_n,\z_n,\ldots,\z_1.
\end{eqnarray*}

In the rest of this subsection
we will examine in detail the one-point function,
\begin{eqnarray*}
{\cal M}^{(i)}(\z;r)&=&
\frac{{}_B\bra{i}\left(E_{++}^{(i)}(\z)-E_{--}^{(i)}(\z)\right)\ket{i}_B}
{{}_B\langle i|i\rangle_B}\cr
&=&g\Bigl(P^{(i)}_{-+}(-q^{-1}\z,\z)-P^{(i)}_{+-}(-q^{-1}\z,\z)\Bigr).
\end{eqnarray*}
By virtue of the formulas (\ref{eqn:eigVO}), (\ref{eqn:leigVO}) and
(\ref{eqn:bunitarity}),
${\cal M}^{(i)}(\z;r)$ has the symmetry
\be\label{eqn:1ptsym}
{\cal M}^{(i)}(\z^{-1};r)={\cal M}^{(i)}(\z;r).
\en
{}From (\ref{eqn:Symm}) we have in addition
\begin{equation}\label{eqn:1ptsym2}
{\cal M}^{(1)}(\z;r)=-{\cal M}^{(0)}(\z;r^{-1}),
\end{equation}
which we shall check directly (see  (\ref{eqn:symm1pt}) below).

Specializing the integral formula (\ref{eqn:Npt}) to
$N=2$, $\z_1=-q^{-1}\z$ and $\z_2=\z$, we obtain
\be
gP^{(i)}_{\mp\pm}(-q^{-1}\z,\z)
=\mp g(z)p^{(i)}_\pm(z,r)
\label{eqn:Epmpm}
\en
where
\begin{eqnarray*}
&&g(z)=\frac{(q^2;q^2)^{4}_\infty}{(q^4;q^4)^{2}_\infty}
\frac{\Theta_{q^4}(z^2)}{1-z^2}=g(z^{-1}),
\\
&&p^{(i)}_\pm(z,r)=
\ \oint_{C^{(i)}_{\pm\pm}}
{dw\over 2\pi \sqrt{ -1}}\ F(w,z,r),
\\
&&F(w,z,r)=
\frac{(1-rz)(1-q^{-2}wz)}{z(1-q^{-2}wr)}
\frac{\Theta_{q^4}(q^2w^2)}{\Theta_{q^2}(wz)\Theta_{q^2}(w/z)}.
\end{eqnarray*}
The contour $C^{(0)}_{--}$ encircles the points $q^{2l}z$,
$q^{2l}z^{-1}$ with $l=1,2,\ldots$, but not the point $q^2r^{-1}$. The
contour $C^{(0)}_{++}$ encircles $q^{2l}z$, $q^{2l}z^{-1}$ with
$l=2,3,\ldots$, but not $q^2r^{-1}$.  (There is no pole at
$w=q^2z^{-1}$.)  The contour $C^{(1)}_{\pm\pm}$ encircles the point
$q^2r^{-1}$ in addition to the same points as $C^{(0)}_{\pm\pm}$ does.
With the change of variables $w\rightarrow q^4/w$ we find
\bea
&&P^{(1)}_{\vep_1,\vep_2}(-q^{-1}\z,\z)=
P^{(0)}_{-\vep_1,-\vep_2}(-q^{-1}\z,\z)\Bigr|_{r\to r^{-1}},
\label{eqn:symm1pt}\\
&&p^{(0)}_\pm(z,r)=-p^{(0)}_\mp(z^{-1},r^{-1})+G(z,r),
\label{eqn:symp}
\ena
where
\begin{eqnarray*}
G(z,r)&=&-{\rm Res}_{w=q^2r}F(w,z^{-1},r^{-1})
\nonumber\\
&=&(1-rz)(1-r/z)
\frac{\Theta_{q^4}(q^2r^2)}{\Theta_{q^2}(rz)\Theta_{q^2}(r/z)}.
\end{eqnarray*}
Using (\ref{eqn:symp}), (\ref{eqn:1ptsym}),
and noting that $G(z,r)=G(z^{-1},r)$, we have
\begin{eqnarray}
&&-\left({\cal M}^{(0)}(\z;r)+{\cal M}^{(0)}(\z;r^{-1})\right)
\nonumber\\
&&=
g(z)\left(p^{(0)}_+(z^{-1},r)+p^{(0)}_-(z^{-1},r)+p^{(0)}_+(z,r^{-1})
+p^{(0)}_-(z,r^{-1})\right)
\nonumber\\
&&=2g(z)G(z,r).
\label{eqn:M01}
\end{eqnarray}

The integral for $p^{(0)}_\pm(z,r)$
can be evaluated explicitly by the calculus of residues,
making use of the quasi-periodicity property
$\Theta_{p}(pz)=-z^{-1}\Theta_{p}(z)$.
We find
\begin{eqnarray}
-{\cal M}^{(0)}(\z;r)
&=&g(z)\left(p^{(0)}_+(z,r)+p^{(0)}_-(z,r)\right)
\nonumber\\
&=&1+2\sum_{l=1}^{\infty}
\frac{(-q^{2})^l(1-rz)(1-r/z)}{ (1-q^{2l}rz)(1-q^{2l}r/z)}.
\label{eqn:1pt}
\end{eqnarray}
The symmetry (\ref{eqn:1ptsym}) is now manifest in the expression.

The boundary
magnetization ${\cal M}^{(i)}(r)$ in the sector $i$ is
\[
{\cal M}^{(i)}(r)=
\frac{{}_B\bra{i}\sigma^z_1\ket{i}_B}{{}_B\langle i|i\rangle_B}
={\cal M}^{(i)}(1;r).
\]
Noting (\ref{eqn:1ptsym2}), we have
\begin{eqnarray}
&&
-{\cal M}^{(0)}(r)={\cal M}^{(1)}(r^{-1})
=1+2\sum_{l=1}^\infty\frac{(-q^2)^l(1-r)^2}{(1-q^{2l}r)^2},
\label{eqn:Mag}\\
&&
{\cal M}^{(1)}(r)-{\cal M}^{(0)}(r)=
2\frac{(q^2;q^2)^4_{\infty}}{(q^2;q^4)^2_\infty}
\frac{(q^2r^2;q^4)_{\infty}(q^2/r^2;q^4)_\infty}
{(q^2r;q^2)^2_{\infty}(q^2/r;q^2)^2_\infty}.
\label{eqn:Magdif}
\end{eqnarray}

When $r$ takes the values $r_c^{(1)}=-q$ and $r_c^{(2)}=q$,
corresponding to the critical fields $\hca$ and $\hcb$,
the difference (\ref{eqn:Magdif}) is 0, as it should be~\cite{MW}. At
these points, the two boundary magnetizations are equal (at the edge of
the hysteresis loop).

Since the ground state is $\ket{0}_B$ for $h\ge 0$ and $\ket{1}_B$ for
$h\le 0$, the boundary magnetization of the ground state is
\[
{\cal M}(r)=\cases{{\cal M}^{(0)}(r)& for $|r|< 1$, \cr
{\cal M}^{(1)}(r)& for $|r|> 1$. \cr}
\]
We see that at $h=0$ ($r=-1$)
the spontaneous magnetization
${\cal M}^{(0)}(-1)=-{\cal M}^{(1)}(-1)$ is nonvanishing.
Specializing (\ref{eqn:Magdif}) to $r=-1$,
we find
\be
{\cal M}^{(0)}(-1)=
-\frac{(q^2;q^2)^4_{\infty}}{(-q^2;q^2)^4_{\infty}}.
\en
Note that $-{\cal M}^{(0)}(-1)$ is the square of the bulk
magnetization~\cite{baxter}.

One can check the formula (\ref{eqn:1pt}) by comparing
(\ref{eqn:Magdif})
with the derivative of the energy difference $\Delta
e(r)=e^{(1)}(r)-e^{(0)}(r)$
with respect to the field $h$.
One can verify directly, by differentiating (\ref{eqn:e1}), that the
following relation holds:
\be\label{eqn:depp}
\frac{\partial \Delta e(r)}{\partial h}=
{\cal M}^{(1)}(r)-{\cal M}^{(0)}(r).
\en

\setcounter{equation}{0}
\section{Rational limit}

Grisaru et al. \cite{GMN} calculated the boundary $S$-matrix
for the $\XXX$ chain using the Bethe ansatz method.
To make comparison, let us consider the limit $q\rightarrow -1$.
In order to get a non-trivial boundary term, it is necessary to
scale $r$ near the point $r=1$.
Setting
\[
r=(q^2)^{\alpha}
\]
the Hamiltonian (\ref{eqn:HBOUND}) scales to
\[
H_B^{\rm bare}=\frac{1}{2}\sum_{k=1}^{\infty}
\Bigl(-\sigma^x_{k+1}\sigma^x_k-\sigma^y_{k+1}\sigma^y_k
+\sigma^z_{k+1}\sigma^z_k\Bigr)
+h\sigma^z_1~,
\]
where $h=1/2\alpha$.
It turns out that we should also scale the parameter $\xi$
in~(\ref{eqn:Mr}) as it approaches $\xi=\sqrt{-1}$.
If we set
\[
\xi=\sqrt{-1}(q^2)^{\sqrt{-1}\lambda/2},
\]
the formula (\ref{eqn:Mr}) for $M(\xi;r)$ can be expressed in terms of the
$q$-gamma function. For example,
\begin{eqnarray*}
&&M^-_-(\xi;r)=(q^2)^{-\sqrt{-1}\lambda}
\\
&&\times
\frac{\Gamma_{q^4}\left(\frac{1}{4}+\frac{\alpha+\sqrt{-1}\lambda}{2}\right)}
{\Gamma_{q^4}\left(\frac{1}{4}+\frac{\alpha-\sqrt{-1}\lambda}{2}\right)}
\frac{\Gamma_{q^4}\left(\frac{3}{4}+\frac{\alpha-\sqrt{-1}\lambda}{2}\right)}
{\Gamma_{q^4}\left(\frac{3}{4}+\frac{\alpha+\sqrt{-1}\lambda}{2}\right)}
\frac{\Gamma_{q^8}\left(\frac{1}{4}-\frac{\sqrt{-1}\lambda}{2}\right)}
{\Gamma_{q^8}\left(\frac{1}{4}+\frac{\sqrt{-1}\lambda}{2}\right)}
\frac{\Gamma_{q^8}\left(1+\frac{\sqrt{-1}\lambda}{2}\right)}
{\Gamma_{q^8}\left(1-\frac{\sqrt{-1}\lambda}{2}\right)}.
\end{eqnarray*}

In the limit $q\rightarrow-1$, the $q$-gamma function is the
ordinary gamma function, and
$$
\frac{M_+^+(\lambda;\alpha)}{M_-^-(\lambda;\alpha)} =
\frac{\sqrt{-1}\lambda + \alpha - \frac{1}{2}}
{\sqrt{-1}\lambda - \alpha + \frac{1}{2}}.
$$
Denoting by ${\cal R}(\lambda)$ and ${\cal K}(\lambda;\alpha)$
the $R$- and $K$-matrices in \cite{GMN}
(see eq. (5.17), (5.19), (5.30) and (5.36) in \cite{GMN}),
\begin{eqnarray*}
\lim\,S_{12}(\xi_1/\xi_2) &=& -\sigma^z_1{\cal R}(\lambda_1-\lambda_2)
\sigma^z_2,
\\
\lim\,M(\xi;r) &=& -\sigma^z{\cal K}(\lambda;\alpha).
\end{eqnarray*}

Following \cite{Smbk}, p.99, let us modify the signs and
define the limiting states by
\begin{equation}\label{eqn:limst}
\ket{\lambda_n,\cdots,\lambda_1}_{\mu_n,\cdots,\mu_1;(i)B}
=\prod_{j:{\rm odd}}(-\mu_j)\lim\,
\Psi^*_{\mu_n}(\xi_n)\cdots\Psi^*_{\mu_1}(\xi_1)\ket{i}_B
\end{equation}
where $\xi_j=\sqrt{-1}(q^2)^{\sqrt{-1}\lambda_j/2}$.
Notice that
\[
\ket{-\lambda_1}_{\mu_1;(i)B}=
\lim \,\Psi^*_{\mu_1}(-\xi_1^{-1})\ket{i}_B
=(-1)^{i+(1-\mu)/2}\lim\, \Psi^*_{\mu_1}(\xi_1^{-1})\ket{i}_B.
\]
Taking into account the signs properly,
we find that in the basis (\ref{eqn:limst})
our bulk and boundary $S$-matrices become in the limit
$-{\cal R}(\lambda_1-\lambda_2)$ and
$-{\cal K}(\lambda;\alpha)$, respectively.

Thus, in the $\XXX$ model, there is only one critical field, which is at
$\alpha=1/2$,
and it is the limit of our $\hcb$. The pole
in the boundary $S$-matrix at $\lambda = \sqrt{-1}(\frac{1}{2}-\alpha)$
is in the physical strip, $0<{\rm Im}\,\lambda\le 1$, when $h>\hcb=1$.
The energy difference and the boundary magnetization in this limit are:
\begin{eqnarray*}
\lim\,\Delta e(r)&=& \frac{\pi}{\sin \pi\alpha},
\\
\lim\,{\cal M}^{(0)}(r)
&=&-1-2\sum_{l=1}^\infty(-1)^l\frac{\alpha^2}{(l+\alpha)^2}
\\
&=&1+\frac{\alpha^2}{2}
\left(\psi'\Bigl(\frac{\alpha+1}{2}\Bigr)
-\psi'\Bigl(\frac{\alpha}{2}\Bigr)\right)
\end{eqnarray*}
where
$\psi'(x)=(d^2/dx^2)\log \Gamma(x)$.

\setcounter{equation}{0}
\section{Discussion}

In conclusion, we have obtained the following results
for the $\XXZ$ Hamiltonian on the semi-infinite chain (\ref{eqn:HBOUND}):
(i) the space of states and the renormalized transfer matrix (\ref{eqn:TB}),
(ii) the vacuum vectors (\ref{eqn:Fi}) and the energy difference
between them (\ref{eqn:e1}),
(iii) the boundary S-matrix (\ref{eqn:Mr}),
and
(iv) the correlation functions (\ref{eqn:Npt}) and
the boundary magnetization (\ref{eqn:Mag}) in particular.
We have also found the critical values of the boundary magnetic field
(\ref{eqn:hcrit}) at which one of the metastable vacuum states
turn to be unphysical. Above the second critical field, one of the
vacuum states has an energy higher than the mass gap.

Contrary to the case of the infinite chain, the symmetry algebra
corresponding to the boundary problem is not $\uq$, since the boundary
Hamiltonian does not commute with it. In this paper, we have bypassed
this point to get the results.  Sklyanin has suggested to us that the
relevant algebra may be the one associated with the boundary YBE as
formulated in his paper \cite{Skl}. In order to discuss highest weight
representations for this algebra, we would probably need a central
extension, just as was discussed by Reshetikhin and
Semenov-Tian-Shansky \cite{RST} in the case of the quantum affine
algebras.

Our approach allows us to diagonalize the transfer
matrix~\refeq{TBfin} on the semi-infinite lattice. However, the
corresponding six--vertex model contains alternating spectral
parameters (see Figure 2).  Therefore the corner transfer matrix is
not diagonalized by our vertex operators, since in
\cite{DFJMN,JMMN,JM} a uniform spectral parameter is assumed.  In
particular, we have not given a method of diagonalizing the transfer
matrix acting in the transverse direction. This would involve
reformulation of the vertex operators to accommodate nonuniform
spectral parameters.

In the field theory limit, there are also two different directions in
which we can formulate the Hamiltonians, corresponding to the exchange
of the space and time coordinates. Sklyanin's transfer matrix
corresponds to a spatial boundary condition. If the roles of space and
time are exchanged, the spatial boundary condition is replaced by some
initial condition in time.  In \cite{GZ} this is represented by the
`boundary state' $|B\rangle$.  Such a state has also been discussed in
the context of conformal field theory~\cite{Ish,CKLM}. Our
expression~(\ref{eqn:grst}) for $|i\rangle_B$ bears some resemblance
to the expression for $|B\rangle$, e.g. (3.38) of~\cite{GZ}.  However,
the two states exist in different spaces. The states $|i\rangle_B$
correspond (see (\ref{eqn:bdst2})) to the `boundary bound states' of~\cite{GZ}.

Although there exists a Bethe ansatz for the
model~\cite{Gaudin,Alc,Skl}, the excitation structure have not yet
been analyzed using this method. In the case of periodic boundary
conditions, only even-number particle states appear in the spectrum of
the Hamiltonian, although both even and odd states are found using the
vertex operator approach. It is unclear whether all of the states we
describe appear in the Bethe ansatz for the boundary problem.  The
boundary scattering matrix (\ref{eqn:Mr}) should also be obtainable
using the Bethe ansatz.

We believe that the case of a non-diagonal $K$-matrix, corresponding
to a boundary magnetic field acting on all three spin components, can
be analyzed using a similar method to the one described in this paper.
Finally, one may ask whether there exist difference equations for
correlation functions of the boundary Hamiltonian, as for the bulk case.

\vskip 1cm
\noindent{\sl Acknowledgements.\quad}
We wish to thank
E.~Corrigan,
O.~Foda,
T.~Inami,
V.~E.~Korepin,
E.~K.~Sklyanin,
P.~B.~Wiegmann
and
A.~B.~Zamolodchikov
for useful discussions.
We are indebted to
B.~M.~McCoy and
R.~I.~Nepomechie for their valuable comments made after reading the
first draft of the manuscript.
We are also grateful to P.~I.~Etingof for sending us his manuscript prior
to publication.
This work is partly supported by Grant-in-Aid for Scientific Research
on Priority Areas 231, the Ministry of Education, Science and Culture.
R.~K.~is supported by the Japan Society for the Promotion of Science.
H.~K.~is supported by Soryushi Shyogakukai.

\begin{appendix}

\setcounter{equation}{0}
\section{Vertex operators}
We summarize here some basic formulas for the
$R$-matrix and the vertex operators, following
the Appendix in \cite{JM}.

\subsection{$R$-matrix}

The $R$-matrix for the six-vertex model is
\begin{equation}\label{eqn:R}
R(\z)=\frac{1}{\kappa(\z)}
\left(\matrix{
1  &         &          &    \cr
&\disp{\frac{(1-\z^2)q}{1-q^2\z^2}}&\disp{\frac{(1-q^2)\z}{1-q^2\z^2}}& \cr
&\disp{\frac{(1-q^2)\z}{1-q^2\z^2}}&\disp{\frac{(1-\z^2)q}{1-q^2\z^2}}& \cr
 &          &          & 1 \cr}\right),
\end{equation}
where
\begin{equation}\label{eqn:kappa}
\kappa(\z)=\z{\pr{q^4\z^2;q^4}\pr{q^2\z^{-2};q^4} \over
\pr{q^4\z^{-2};q^4}\pr{q^2\z^{2};q^4}},
\qquad (z;p)_\infty=\prod_{n=0}^\infty(1-zp^n).
\end{equation}
Let $\{v_+,v_-\}$ denote the natural basis of $V=\C^2$.
When viewed as an operator on $V\otimes V$, the matrix elements of $R(\z)$
are defined by
\[
R(\z)\left(v_{\vep'_1}\otimes v_{\vep'_2}\right)=\sum_{\vep_1,\vep_2}
v_{\vep_1}\otimes v_{\vep_2}\,R^{\vep_1'\,\vep_2'}_{\vep_1\,\vep_2}(\z).
\]
As usual, when copies $V_j$ of $V$ are involved,
$R_{ij}(\z)$ acts as $R(\z)$ on the $i$-th and $j$-th
components and as the identity elsewhere.
The Yang-Baxter equation satisfied by (\ref{eqn:R}) is
\begin{equation}\label{YBE}
R_{12}(\z_1/\z_2)R_{13}(\z_1/\z_3)R_{23}(\z_2/\z_3)
=
R_{23}(\z_2/\z_3)R_{13}(\z_1/\z_3)R_{12}(\z_1/\z_2).
\end{equation}
The scalar factor $\kappa(\z)$ (\ref{eqn:kappa}) is so chosen that
the unitarity and crossing relations are
\begin{eqnarray}
&&R_{12}(\z_1/\z_2)R_{21}(\z_2/\z_1)=1,
\label{eqn:unitarity} \\
&&
R^{\vep'_2\vep_1}_{\vep_2\vep_1'}(\z_2/\z_1)
=
R^{-\vep_1'\,\vep_2'}_{-\vep_1\,\vep_2}\bigl(-q^{-1}\z_1/\z_2\bigr).
\label{eqn:crossing}
\end{eqnarray}

\subsection{Vertex operators}

We list here the commutation relations for the vertex operators:
\begin{eqnarray}
&&
\Phi_{\vep_2}(\z_2)\Phi_{\vep_1}(\z_1)=
\sum_{\vep_1',\vep_2'}
R^{\vep_1'\,\vep_2'}_{\vep_1\,\vep_2}(\z_1/\z_2)
\Phi_{\vep_1'}(\z_1)\Phi_{\vep_2'}(\z_2),
\label{eqn:comI}\\
&&\Psi_{\mu_1'}^*(\xi_1)\Psi_{\mu_2'}^*(\xi_2)
=-\sum_{\mu_1,\mu_2}R^{\mu_1'\,\mu_2'}_{\mu_1\,\mu_2}(\xi_1/\xi_2)
\Psi_{\mu_2}^*(\xi_2)\Psi_{\mu_1}^*(\xi_1),
\label{eqn:psicom}\\
&&\Phi_{\vep}(\z)\Psi_{\mu}^*(\xi)=
\tau(\z/\xi)\Psi_{\mu}^*(\xi)\Phi_{\vep}(\z).
\label{eqn:phipsi}
\end{eqnarray}
Here
\begin{equation}\label{eqn:tau}
\tau(\z)=
\z^{-1}{\Theta_{q^4}(q\z^{2}) \over \Theta_{q^4}(q\z^{-2})},
\qquad
\Theta_p(z)=(z;p)_\infty(pz^{-1};p)_\infty (p;p)_\infty.
\end{equation}

The type I vertex operators satisfy the invertibility
\begin{equation}\label{eqn:invert}
g\,\sum_\vep\Phi^*_\vep(\z)\Phi_\vep(\z)=\id,
\qquad
g\,\Phi_{\vep_1}(\z)\Phi^*_{\vep_2}(\z)=\delta_{\vep_1\vep_2}\id,
\end{equation}
where $g$ is given by
\begin{equation}\label{eqn:G}
g=\frac{\pr{q^2;q^4}}{\pr{q^4;q^4}}.
\end{equation}
For the type II operators the corresponding property is
\begin{equation}\label{eqn:delta}
\Psi^{(i,1-i)}_{\mu_1}(\xi_1)
\Psi^{*(1-i,i)}_{\mu_2}(\xi_2)=
{g\delta_{\mu_1\mu_2}\over 1-\xi_2^2/\xi_1^2}
\left({\xi_2\over\xi_1}\right)^{i+{1-\mu_1\over2}}+ \cdots
\qquad (\xi_1\rightarrow \pm\xi_2),
\end{equation}
where
\[
\Psi^{(i,1-i)}_{\mu}(\xi)=
\Psi^{*(i,1-i)}_{-\mu}(-q^{-1}\xi)
\]
and $\cdots$ means regular terms.

\subsection{Bosonization}
For $i=0,1$, consider the bosonic Fock space
\[
\H^{(i)}=\C[a_{-1},a_{-2},\cdots]
\otimes \left(\oplus_{n\in\Z}\C e^{\Lambda_i+n\alpha}\right).
\]
The commutation relations of $a_n$ are
\[
[a_m,a_n]=\delta_{m+n,0}{[m][2m]\over m}~,
\quad (m,n\neq 0),\qquad [n]=\frac{q^n-q^{-n}}{q-q^{-1}}.
\]
On the symbols $e^\gamma$,
the operators $e^\beta$, $z^\partial$ act as
\[
e^\beta . e^\gamma=e^{\beta+\gamma},
\qquad
z^{\partial}.e^\gamma=z^{[\partial,\gamma]}e^\gamma,
\]
where
$[\partial,\alpha]=2$, $[\partial, \Lambda_0]=0$
and $\Lambda_1=\Lambda_0+\alpha/2$.
The highest weight vector of $\H^{(i)}$ is given by
$\ket{i}=1\otimes e^{\Lambda_i}$.

We have the following bosonic realization for the vertex operators:
\begin{eqnarray}
\Phi^{(1-i,i)}_-(\z)
&=&
e^{P(\z^2)}e^{Q(\z^2)}\otimes
e^{\alpha/2}(-q^3\z^2)^{(\partial +i)/2}\z^{-i},
\label{eqn:phim}\\
\Phi^{(1-i,i)}_+(\z)&=&
\oint_{C_1} {dw \over 2\pi i}
{(1-q^2)w\z \over q(w-q^2\z^2)(w-q^4\z^2)}:\Phi^{(1-i,i)}_-(\z)X^-(w):,
\nonumber\\
&&\label{eqn:phip}\\
\Psi^{*(1-i,i)}_-(\z)&=&
e^{-P(q^{-1}\z^2)}e^{-Q(q\z^2)}\otimes
e^{-\alpha/2}(-q^3\z^2)^{(-\partial +i)/2}\z^{1-i},
\label{eqn:psim}\\
\Psi^{*(1-i,i)}_+(\z)&=&
\oint_{C_2} {dw \over 2\pi i}
{q^2(1-q^2)\z \over (w-q^2\z^2)(w-q^4\z^2)}:\Psi^{*(1-i,i)}_-(\z)X^+(w):,
\nonumber\\
&&\label{eqn:psip}\\
X^{\pm}(z)&=&e^{R^\pm(z)}e^{S^\pm(z)}\otimes e^{\pm \alpha}z^{\pm\partial},
\label{eqn:xpm}
\end{eqnarray}
where
\begin{eqnarray}
&&P(z)=\sum_{n=1}^\infty \disp{\frac{a_{-n}}{[2n]}}q^{7n/2}z^n,
\qquad
Q(z)=-\sum_{n=1}^\infty \disp{\frac{a_{n}}{[2n]}}q^{-5n/2}z^{-n},
\label{eqn:PQ}\\
&&R^{\pm}(z)=
\pm \sum_{n=1}^\infty \disp{\frac{a_{-n}}{[n]}}q^{\mp n/2}z^n,
\quad
S^\pm(z)=\mp \sum_{n=1}^\infty \disp{\frac{a_{n}}{[n]}}q^{\mp n/2}z^{-n}.
\label{eqn:RS}
\end{eqnarray}
The integration contours encircle $w=0$ in such a way that
\begin{eqnarray*}
C_1&:&\hbox{ $q^4\z^2$ is inside and $q^2\z^2$ is outside}, \\
C_2&:&\hbox{ $q^4\z^2$ is outside and $q^2\z^2$ is inside}.
\end{eqnarray*}

These operators act also on
the dual (right) modules $\H^{*(i)}$,
which are defined analogously by replacing $e^{\Lambda_i+n\alpha}$
with $e^{-(\Lambda_i+n\alpha)}$.
In particular the highest weight vector
is $\bra{i}=1\otimes e^{-\Lambda_i}$.
The right actions of $e^\beta$ and $z^{\partial}$ are given by
\[
e^\gamma . e^\beta=e^{\gamma+\beta},
\qquad
e^\gamma . z^\partial = e^\gamma z^{[\gamma,\partial]}.
\]

\setcounter{equation}{0}

\section{The $q$-expansion}
In this section we calculate the ground state vector $|0\rangle_B$
of the Hamiltonian (\ref{RENOR}),
as a $q$-series expansion in terms of paths for $r=-1$
\cite{FM,DFJMN}.
Next we develop the vector $\ket{0}_B$
written in terms of bosonic operators into a similar $q$-series.
We find that the two results agree to the order $q^3$.

\subsection{The path expansion}
In the na{\"\i}ve approach,
we can expand the ground state vector $|0\rangle_B$ as a linear combination
of vectors $|p\rangle$ in the semi--infinite tensor product
$\otimes_{k\ge1}V_k$ (see \refeq{BCI}):
\[
|p\rangle=\otimes_{k\ge1}v_{p(k)},
\]
where
\[
p:\Z_{\ge1}\rightarrow\{\pm\},\quad p(k)=(-1)^k \hbox{ if }k\gg 1.
\]
We call such a map $p$ a path, as well as the vector $|p\rangle$ itself.
In what follows, we find it convenient to represent a path $p$ by
a sequence of non-negative integers
\[
Y=(f_1,f_2,\cdots,f_n), \qquad f_1>f_2>\cdots>f_n,
\]
given via the correspondence
\[
p(k)=p(k+1)\hbox{ if and only if }k\in Y.
\]
We also use the convention that
$|f_1,\cdots,f_n,0\rangle=|f_1,\cdots,f_n\rangle$.
The degree of $Y$ is  $f_1+\cdots+ f_n$.
The `vacuum path' $p_0$ given by $p_0(k)=(-1)^k$
is the unique one of degree $0$, which we represent by the symbol $Y=\phi$.

Let
\be\lb{APPL}
|0\rangle_B=|\phi\rangle+\sum_{j\ge1}\vep^jv_j
\en
be the expansion of the ground state vector for (\ref{eqn:HBOUND}), where
$\vep=-1/(2\Delta)=-q/(1+q^2)$.
Each $v_j$ is a linear combination of paths
\be
v_j=\sum_{Y\ne \phi}c_j(Y)|Y\rangle.
\en
Here we demanded that $\phi$ never appears in $v_j$ ($j\ge 1$),
by multiplying a scalar to $\ket{0}_B$ as necessary.

We are going to solve the following eigenvalue
equation order by order in $\vep$:
\[
\left(\sum_{k\ge1}{1\over2}(\sigma^z_{k+1}\sigma^z_k+1+c_k(\vep))
-2\vep Q \right)|0\rangle_B=0,
\]
where
\[
Q=\sum_{k\ge1}(\sigma^+_{k+1}\sigma^-_k+\sigma^-_{k+1}\sigma^+_k).
\]
We have included the $c$-number terms
$c_k(\vep)=\sum_{j\ge1}\vep^jc_{k,j}$
to ensure that the eigenvalue is 0.

Up to the third order, we obtain
\begin{eqnarray}
&&v_1=2|2\rangle+\sum_{k\ge1}|k+2,k\rangle,
\label{eqn:v1}\\
&&v_2=6|4\rangle+2\sum_{k\ge1}|k+4,k\rangle
+2\sum_{k\ge3}|k+2,k,2\rangle
+\sum_{k\ge1\atop l\ge k+3}|l+2,l,k+2,k\rangle,
\nonumber\\
&&\label{eqn:v2}\\
&&v_3=-4|2\rangle+20|6\rangle-3|3,1\rangle+2|4,2\rangle
-\sum_{k\ge3}|k+2,k\rangle+5\sum_{k\ge1}|k+6,k\rangle\nonumber\\
&&+4|4,3,1\rangle+4\sum_{k\ge3}|k+4,k,2\rangle
+6\sum_{k\ge5}|k+2,k,4\rangle
+\sum_{k\ge1}|k+4,k+3,k+1,k\rangle\nonumber\\
&&+2\sum_{k\ge1\atop l\ge k+5}|l+2,l,k+4,k\rangle
+2\sum_{k\ge1\atop l\ge k+3}|l+4,l,k+2,k\rangle\nonumber\\
&&+2\sum_{k\ge3\atop l\ge k+3}|l+2,l,k+2,k,2\rangle
+\sum_{k\ge1\atop{l\ge k+3\atop m\ge l+3}}|m+2,m,l+2,l,k+2,k\rangle.
\nonumber\\
&&\label{eqn:v3}
\end{eqnarray}

\subsection{Monomials in bosons and the upper global bases}
Next, we expand bosonic expression for the vector $\ket{0}_B$
(\ref{eqn:Fi}) in terms of paths.  Since $\ket{0}_B$ is written in
terms of bosons, there is a difficulty in handling its expansion in
the sense of crystal lattice. We expect that each homogeneous
component of $\ket{0}_B$ belongs to the upper crystal lattice (see
\cite{DFJMN}), although we have no proof.  On the other hand, the
global basis vectors have a $q$-expansion in terms of paths. (For the
meaning of this statement, see \cite{FM,DFJMN}, in particular, the
remark at the end of page 99 of \cite{DFJMN} and \cite{Etingof}.)  In
the following we shall relate the vectors created by bosons to the
global base vectors up to degree $6$. This will make it possible (to
this degree) to develop the homogeneous components of $\ket{0}_B$ into
a
$q$-series, which can be compared to the results of the previous
subsection.

For a path $Y$, denote by $G(Y)\in\H^{(0)}$
the corresponding upper global base of Kashiwara
(see Section 2 of \cite{DFJMN}).
Up to degree $6$, the possible paths and monomials in the bosons
are as follows.

\medskip
\def\arraystretch{1.5}
\begin{center}
  \begin{tabular}{|c|c|c|}\hline
\hbox{degree}&\hbox{path}&\hbox{monomials}\\ \hline
$0$&$\phi$& $1$\\ \hline
$1$&($1$)& $e^\alpha$ \\ \hline
$2$&($2$)& $a_{-1}$ \\ \hline
$3$&($3$),($2,1$)& $a_{-1}e^\alpha$, $e^{-\alpha}$\\ \hline
$4$&($4$),($3,1$)&$a_{-1}^2$, $a_{-2}$\\ \hline
$5$&($5$),($4,1$),($3,2$)&$a_{-1}e^{-\alpha}$, $a_{-1}^2e^\alpha$,
$a_{-2}e^\alpha$\\ \hline
$6$&($6$),($5,1$),($4,2$),($3,2,1$)
&$a_{-3}$, $a_{-1}a_{-2}$, $a_{-1}^3$, $e^{2\alpha}$\\ \hline
\end{tabular}
\end{center}
In the second column
we abbreviated $X\otimes e^{\Lambda_0+\beta}$ to $Xe^\beta$.

The following is the explicit relation between the two bases:
\begin{eqnarray*}
&&G(\phi)=1\otimes e^{\L_0},\\
&&G(1)=q^2\otimes e^{\L_0+\alpha},\\
&&G(2)=-{q^{5/2}\over[2]}a_{-1}\otimes e^{\L_0},\\
&&G(3)={q^{5/2}\over[2]}a_{-1}\otimes e^{\L_0+\alpha},\\
&&G(2,1)=-q^3\otimes e^{\L_0-\alpha}\\
&&G(4)=\Bigl({q^5\over2[2]^2}a_{-1}^2-{q^5\over[4]}a_{-2}\Bigr)
\otimes e^{\L_0},\\
&&G(3,1)=\Bigl(-{q^3\over2[2]}a_{-1}^2+{q^4-q^2\over[4]}a_{-2}\Bigr)
\otimes e^{\L_0},\\
&&G(5)=\Bigl({q^3\over2[2]^2}a_{-1}^2+{q^3\over[4]}a_{-2}\Bigr)
\otimes e^{\L_0+\alpha},\\
&&G(4,1)={q^{11/2}\over[2]}a_{-1}\otimes e^{\L_0-\alpha},\\
&&G(3,2)=\Bigl({q^3\over2[2]^2}a_{-1}^2-{q^3\over[4]}a_{-2}\Bigr)
\otimes e^{\L_0+\alpha},\\
&&G(6)=q^{5/2}\Bigl(-{q^5\over[6]}a_{-3}+{q^5\over[4][2]}a_{-2}a_{-1}
-{q^5\over6[2]^3}a_{-1}^3\Bigr)\otimes e^{\L_0},\\
&&G(5,1)=q^{5/2}\Bigl({q^4-1\over[6]}a_{-3}-{q^2\over[2]^2}a_{-2}a_{-1}
+{q(q^2-1)\over6[2]^2}a_{-1}^3\Bigr)\otimes e^{\L_0},\\
&&G(4,2)=q^{5/2}\Bigl(-{q(q^2-1)\over[6]}a_{-3}+{q\over[4][2]}a_{-2}a_{-1}
+{q(2q^2+1)\over6[2]^3}a_{-1}^3\Bigr)\otimes e^{\L_0},\\
&&G(3,2,1)=q^6\otimes e^{\Lambda_0+2\alpha}.
\end{eqnarray*}

{}From these data, we can write $|0\rangle_B$  (see \refeq{Fi})
as a linear combination of the global basis vectors up to degree $6$:
\begin{eqnarray}
|0\rangle_B&=&G(\phi)+qrG(2)+q^2r^2G(4)+q^3G(3,1)+q^3r^3G(6)
\nonumber\\
&&
+q^4[2]rG(4,2)+q^4rG(5,1)+\cdots.
\lb{APP1}
\end{eqnarray}
The global base vectors are regular and finite at $q=0$.
We say $\ket{0}_B$ is regular to mean that each coefficient
in the expansion (\ref{eqn:APP1}),
regarded as a function of $q$ by setting  $r=\pm (q^2)^\alpha$,
is regular at $q=0$.
Hence we find, up to this order, that this is the case if
and only if $|rq|\le1$.

Now, we expand $|0\rangle_B$ in $q$.
Let
\[
G(Y)=|p_0(Y)\rangle+q|p_1(Y)\rangle+q^2|p_2(Y)\rangle+\cdots,
\]
be the expansion of a global base vector $G(Y)$,
where $|p_j(Y)\rangle$ is a finite or possibly infinite linear combination
of paths
\[
|p_j(Y)\rangle=\sum_{Y'}c_j(Y',Y)|Y'\rangle.
\]
We have, in particular,
\begin{equation}\label{eqn:APP2}
c_0(Y',Y)=\delta_{Y',Y}.
\end{equation}

We have the following expansions.
\bea
&&G(\phi)=|\phi\rangle
+\vep\sum_{k\ge0}|k+2,k\rangle
+\vep^2\Bigl(2\sum_{k\ge0}|k+4,k\rangle
+\sum_{l\ge0\atop k\ge l+3}|k+2,k,l+2,l\rangle\Bigr)\nonumber\\
&&+\vep^3\Bigl(\sum_{m\ge0\atop{l\ge m+3\atop k\ge l+3}}
|k+2,k,l+2,l,m+2,m\rangle
+2\sum_{k\ge0\atop l\ge k+5}|l+2,l,k+4,k\rangle\nonumber\\
&&+2\sum_{k\ge0\atop l\ge k+3}|l+4,l,k+2,k\rangle
+\sum_{k\ge0}|k+4,k+3,k+1,k\rangle
+5\sum_{k\ge0}|k+6,k\rangle\nonumber\\
&&-\sum_{k\ge1}|k+2,k\rangle\Bigr)+O(\vep^4),\lb{APP3}\\
&&G(2)=(1-4\varepsilon^2+O(\varepsilon^4))\Bigl\{
|2\rangle+\vep\Bigl(-|0\rangle+3|4\rangle+
\sum_{k\ge3}|k+2,k,2\rangle\Bigr)\nonumber\\
&&+\vep^2\Bigl(9|6\rangle-2|3,1\rangle-\sum_{k\ge3}|k+2,k\rangle
+2|4,3,1\rangle+3\sum_{k\ge5}|k+2,k,4\rangle\nonumber\\
&&+2\sum_{k\ge3}|k+4,k,2\rangle
+\sum_{k\ge3\atop l\ge k+3}|l+2,l,k+2,k,2\rangle\Bigr)
+O(\vep^3)\Bigr\},\lb{APP4}\\
&&G(4)=|4\rangle
+\vep\Bigl(-3|2\rangle
+5|6\rangle+|4,2\rangle+|4,3,1\rangle
+\sum_{k\ge5}|k+2,k,4\rangle\Bigr)+O(\vep^2).\nonumber\\
\lb{APP5}
\ena

Combining \refeq{APP1}--\refeq{APP5},
we obtain the $q$-expansion of $|0\rangle_B$. We have checked the equality
between \refeq{APP1} for $r=-1$ and
\refeq{APPL}, (\ref{eqn:v1})--(\ref{eqn:v3}) up to the third order.

\setcounter{equation}{0}
\section{Coherent states}
We here summarize formulas concerning coherent states of
bosons which are used in Section 4.

The coherent states $\ket{ \xi}_i$ and ${}_i\bra{{\bar\xi}}$ in the Fock
spaces ${\cal H}^{(i)},{\cal H}^{*(i)}, \ i=0, 1$, are defined by
\bea
\ket{ \xi}_i&=&\exp\Bigl({\sum_{n=1}^\infty
{n\over [n][2n]}\xi_na_{-n}}\Bigr)\ket{i},\\
{}_i\bra{\bar\xi}&=&\bra{i}
\exp\Bigl({\sum_{n=1}^\infty {n\over [n][2n]}{\bar\xi_n}a_{n}}\Bigr),
\ena
where $\xi_n$ and ${\bar \xi}_n$ are complex conjugate parameters.

Noting that the highest weight states $\ket{i}$ and $\bra{i}$ are
annihilated by the
boson oscillators $a_n$ and $a_{-n}$ with $n \geq 1$, respectively, one can
easily verify
\be
a_n\ket{\xi}_i=\xi_n\ket{\xi}_i, \qquad
{}_i\bra{{\bar\xi}}a_{-n}={\bar\xi_n}~{}_i\bra{{\bar\xi}}.
\en

One can also show that the coherent states $\{\ket{\xi}_i\}$
(resp. $\{{}_i\bra{{\bar \xi}}\}$) form a complete basis in the Fock
space
${\cal H}^{(i)}$ (resp.\ ${\cal H}^{*(i)} )$. Namely one can verify
the completeness relation
\be
{\rm id}_{{\cal H}^{(i)}}=\displaystyle{
                  \int \prod_{n=1}^{\infty}
\frac{nd\xi_nd{\bar\xi_n}}{[n][2n]}~
                  e^{-\sum_{n=1}^{\infty}\frac{n}{[n][2n]}|\xi_n|^2}
                  \ket{\xi_n}_i{}_i\bra{\bar{\xi_n}}}.
\label{eqn:comp}
\en
Here the integration is taken over the entire complex plane with the
measure $d\xi d{\bar\xi}=dxdy$ for $\xi=x+iy$.
In the proof, the following integration formula is used:
\bea
\lefteqn{
\displaystyle{\int \prod_{n=1}^{\infty}
\frac{nd\xi_nd{\bar\xi_n}}{[n][2n]}~
\exp\Bigl(
-{1\over2}\sum_{n=1}^{\infty}\frac{n}{[n][2n]}
({\bar\xi}_n,\xi_n){\cal A}_n\left(\matrix{{\bar\xi}_n\cr
\xi_n}\right)
+\sum_{n=1}^\infty({\bar\xi}_n,\xi_n)\ {\cal B}_n
\Bigr)
}}\cr
&&=\prod_{n=1}^\infty\Bigl(-\det{\cal A}_n\Bigr)^{-1/2}
\exp\Bigl(\frac{1}{2}\sum_{n=1}^\infty\frac{[n][2n]}{n}
{\cal B}^t_n{\cal A}^{-1}_n{\cal B}_n\Bigr),\qquad\qquad\qquad
\label{eqn:Cohint}
\ena
where ${\cal A}_n$ are invertible constant 2$\times$2 matrices
and ${\cal B}_n $ are constant 2 component vectors.

\end{appendix}

\bibliographystyle{unsrt}

\begin{thebibliography}{10}

\bibitem{MWa} McCoy, B.~M. and Wu, T.~T.,
\newblock Theory of Toeplitz determinants and the spin correlation
functions of the two-dimensional Ising model. IV,
\newblock {\it Phys. Rev.} {\bf 162} 436--475 (1967).

\bibitem{MWb} McCoy, B.~M. and Wu, T.~T.,
\newblock Theory of Toeplitz determinants and the spin correlation
functions of the two-dimensional Ising model. V,
\newblock {\it Phys. Rev.} {\bf 174} 546--559 (1968).

\bibitem{Bariev}
Bariev, R. Z.,
\newblock Correlation functions of the semi-infinite two-dimensional Ising
model,
\newblock {\it Theoret. Math. Phys.} {\bf 40} 623--626 (1979).

\bibitem{Gaudin}
Gaudin, M.,
\newblock Boundary energy of a bose gas in one dimension,
\newblock {\it Phys. Rev. A} {\bf 4} 386--394 (1971).

\bibitem{Alc}
Alcaraz, F. C., Barber, M. N., Batchelor, M. T., Baxter, R.
and Quispel, G. R. W.,
\newblock Surface exponents of the quantum $\XXZ$, Ashkin-Teller and Potts
models,
\newblock {\it J. Phys. A} {\bf 20} 6397--6409 (1987).


\bibitem{Skl}
Sklyanin, E. K.,
\newblock Boundary conditions for integrable quantum systems,
\newblock {\it J. Phys. A} {\bf 21} 2375--2389 (1988).

\bibitem{KuS}
Kulish, P. P. and Sklyanin, E. K.,
\newblock The general $U_q\left[sl(2)\right]$ invariant $\XXZ$ integrable
quantum spin chain,
\newblock {\it J. Phys. A} {\bf 24} L435--L439 (1991).

\bibitem{MN1}
Mezincescu, L. and Nepomechie, R. I.,
\newblock Integrability of open spin chains with quantum algebra symmetry,
\newblock {\it Int. J. Mod. Phys. A} {\bf 6} 5231--5248 (1991);
\newblock {\it Int. J. Mod. Phys. A} {\bf 7} 5657--55659 (1992)

\bibitem{MN2}
Mezincescu, L. and Nepomechie, R. I.,
\newblock Analytical Bethe ansatz for quantum-algebra-invariant spin chains,
\newblock {\it Nucl. Phys. B} {\bf 372} 597--621 (1992).

\bibitem{GMN}
Grisaru, M. T., Mezincescu, L. and Nepomechie, R. I.,
\newblock Direct calculation of the boundary $S$ matrix for the Heisenberg
chain,
\newblock {\it preprint} {UMTG-177, hep-th/9407089}, to appear in {\it
J. Phys. A}.

\bibitem{dVGR}
de Vega, H. J. and Gonz\'alez Ruiz, A.,
\newblock Boundary $K$-matrices for the six vertex and the
$n(2n-1)$ $A_{n-1}$ vertex models,
\newblock {\it J. Phys. A} {\bf 26} L519--524 (1993).

\bibitem{IK}
Inami, T. and Konno, H.,
\newblock Integrable XYZ spin chain with boundaries,
\newblock {\it preprint} {YITP/K-1084, hep-th/9409138}, to appear in
{\it J. Phys. A. Letts.}.

\bibitem{HSFY}
Hou, B.-Y., Shi, K.-J. Fan, H. and Yang, Z.-X.,
Solution of reflection equation,
preprint NWU-IMP-931201 (1993).

\bibitem{YB}
Yung, C. M. and Batchelor, M. T.,
\newblock Integrable vertex and loop models on the square lattice with open
boundaries via reflection matrices,
\newblock {\it preprint} {MRR042-94, hep-th/9410042} (1994).


\bibitem{Cher}
Cherednik, I. V.,
\newblock Factorizing particles on a half-line and root systems,
\newblock {\it Theor. Math. Phys.} {\bf 61} 977--983 (1984).

\bibitem{GZ}
Ghoshal, S. and Zamolodchikov, A.,
\newblock Boundary $S$-matrix and boundary state in two-dimensional
integrable quantum field theory,
\newblock {\it Int. J. Mod. Phys. A} {\bf 21} 3841--3885 (1994).


\bibitem{Cardy}
Cardy, J. L.,
\newblock Effect of boundary conditions on the operator content of two-
dimensional conformally invariant theories,
\newblock {\it Nucl. Phys.} {\bf B275[FS]} 200--218 (1986).

\bibitem{FS}
Fendley, P. and Saleur, H.,
\newblock Deriving boundary $S$ matrices,
\newblock {\it preprint} {USC-94-001, hep-th/9402045} (1994),
{\it to appear in Nucl. Phys.}

\bibitem{FK}
Fring, A. and K\"oberle, R.,
\newblock Factorized scattering in the presence of reflecting boundaries,
\newblock {\it Nucl. Phys.} {\bf B421} 159--172 (1994).

\bibitem{Sasaki}
Corrigan, E., Dorey, P. E. and Sasaki, R.,
\newblock Affine Toda field theory on a half-line,
\newblock {\it Phys. Lett.} {\bf B333} 83--91 (1994).

\bibitem{DFJMN}
Davies, B., Foda, O., Jimbo, M., Miwa, T. and
Nakayashiki, A.,
\newblock Diagonalization of the $\XXZ$ Hamiltonian by vertex operators,
\newblock {\it Commun. Math. Phys.}
{\bf 151}, 89--153 (1993).

\bibitem{JMMN}
Jimbo, M., Miki, K., Miwa, T. and Nakayashiki, A.,
\newblock Correlation functions of the $\XXZ$ model for $\Delta<-1$,
\newblock {\it Phys. Lett. A}
{\bf 168}, 256--263 (1992).

\bibitem{JM}
Jimbo, M. and  Miwa, T.,
\newblock {\it Algebraic Analysis of Solvable Lattice Models},
\newblock RIMS preprint {\bf 981} (1994),
\newblock {\it to appear in}
CBMS Regional Conference Series in Mathematics, vol.{\bf 85}, AMS.

\bibitem{FJ}
Frenkel, I. B. and Jing, N.,
\newblock Vertex representations of quantum affine algebras,
\newblock Proc. Natl. Acad. Sci. USA, {\bf 85} 9373--9377 (1988).

\bibitem{MW}
McCoy, B. M. and Wu, T. T.,
\newblock {\it The two dimensional Ising model},
\newblock Harvard University Press, Cambridge, Massachusetts,
1973.

\bibitem{PS} Pasquier, V. and Saleur, H.,
\newblock Common structures between finite systems and conformal field
theories through quantum groups,
\newblock {\it Nucl.Phys.} {\bf B330} 523 (1990).

\bibitem{Etingof}
Etingof, P. I.,
\newblock On spectral theory of quantum vertex operators,
\newblock {\it preprint}, hep-th/9410208,
(1994).

\bibitem{baxter}
Baxter, R.~J.,
\newblock Spontaneous staggered polarization of the $F$ model,
\newblock {\it J. Stat. Phys.}, {\bf 9} 145--182 (1973).

\bibitem{Smbk}
Smirnov, F. A.,
\newblock {\it Form Factors in Completely Integrable Models of
Quantum Field Theory},
\newblock Advanced Series in Mathematical Physics {\bf 14},
World Scientific, Singapore 1992.

\bibitem{RST}
Reshetikhin, N. Yu. and Semenov-Tian-Shansky, M. A.,
\newblock Central extensions of quantum current groups,
\newblock {\it Lett. Math. Phys.} {\bf 19} 133--142 (1990).

\bibitem{Ish}
Ishibashi, N.,
The boundary and crosscap states in conformal field theories,
\newblock {\it Mod. Phys. Lett. A} {\bf 4} 251--264 (1989).

\bibitem{CKLM}
Callan, C. G., Klebanov, I. R., Ludwig, A. W. W. and Maldacena, J. M.,
\newblock Exact solution of a boundary conformal  field theory,
\newblock {\it Nucl. Phys.} {\bf  B422 [FS]} 417--448 (1994).

\bibitem{FM}
Foda, O. and Miwa, T.,
\newblock Corner transfer matrices and quantum affine algebras,
\newblock {\it Int. J. Mod. Phys. A} {\bf7} Suppl. 1A,  279--302 (1992).

\end{thebibliography}

\end{document}